\documentclass[pra,twocolumn,showpacs,preprintnumbers,amsmath,amssymb, superscriptaddress,floatfix]{revtex4-2}

\usepackage[makeroom]{cancel}

\usepackage{amsmath}    
\usepackage{amssymb,environ}
\usepackage{graphicx}   
\usepackage{verbatim}   
\usepackage{color}      
\usepackage{url}
\usepackage[normalem]{ulem}
\usepackage{natbib}
\usepackage{enumitem}
\usepackage{multirow}
\usepackage{mathtools}

\usepackage{epsfig}
\usepackage{textcomp}
\usepackage{wasysym}
\usepackage{array}
\usepackage{color}

\begin{document}

\newcommand{\ev}[0]{\mathbf{e}}
\newcommand{\cv}[0]{\mathbf{c}}
\newcommand{\fv}[0]{\mathbf{f}}
\newcommand{\Rv}[0]{\mathbf{R}}
\newcommand{\Tr}[0]{\mathrm{Tr}}
\newcommand{\ud}[0]{\uparrow\downarrow}
\newcommand{\Uv}[0]{\mathbf{U}}
\newcommand{\Iv}[0]{\mathbf{I}}
\newcommand{\Hv}[0]{\mathbf{H}}

\setlength{\jot}{2mm}

\newcommand{\jav}[1]{#1}

\title{\jav{Inelastic tunneling through normal and superconducting junctions in the presence of photonic bath within Lindbladian formalism}}

\author{\'Ad\'am B\'acsi}
\email{bacsi.adam@sze.hu}
\affiliation{Jo\v zef Stefan Institute, Jamova 39, SI-1000 Ljubljana, Slovenia}
\affiliation{Department of Mathematics and Computational Sciences, Sz\'echenyi Istv\'an University, 9026 Gy\H or, Hungary}
\author{Rok \v Zitko}
\affiliation{Jo\v zef Stefan Institute, Jamova 39, SI-1000 Ljubljana, Slovenia}
\affiliation{Faculty of Mathematics and Physics, University of Ljubljana, Jadranska 19, SI-1000 Ljubljana, Slovenia}
\date{\today}

\begin{abstract}
\jav{An electron tunneling across a junction integrated into an electric circuit can generate an excitation in the photonic field (electromagnetic environment) and lose energy in the process. Such inelastic tunneling of particles is commonly described using the $P(E)$ theory.} In the conventional approach to \jav{this} theory, the tunneling rate and the electric current through the junction are derived using Fermi's golden rule \jav{and by} averaging over the environmental photonic degrees of freedom. \jav{In this work, we address the same problem of inelastic tunneling due to photonic environment in Lindbladian formalism and we present how the photonic degrees of freedom are traced out in quantum master equation approach}. 
The resulting quantum master equation \jav{is parameterized by the same $P(E)$ function and} enables us to obtain not only the electric current but various other quantities, for instance the heat current, in a systematic and convenient way. \jav{We also demonstrate that the Lindbladian formalism provides a comprehensive description of Bogoliubov quasiparticle tunneling through superconducting junctions and that it properly accounts for the coherence factors. The coherence factors become important if the normal-state density of states is particle-hole asymmetric.}
\end{abstract}

\maketitle

\section{Introduction}
Tunneling is one of the most peculiar quantum phenomena. In contrast to classical behavior, quantum transmission is also possible through potential barriers exceeding the total energy of the particle. Quantum tunneling has attracted much interest due to the wide range of applications including superconducting quantum interference devices \cite{barone,golubovRMP2004}, scanning tunneling microscopy \cite{RMPSTM1987,chen,RMPtrSTM2010,Ast2016,Karan2022}, atomic scale devices \cite{agrait2003,Bretheau2011,Evers2020}, quantum dots \cite{kouwenhoven1998,wiel2003,hybrid2010}, quantum optical devices \cite{Aiello2022} and resonant tunneling diodes \cite{5391729,Sollner1983}.

In most situations, the tunneling event can be considered a closed (i.e., unitary) process in the sense that the particle does not interact with the environment. In some experiments \cite{likharev1989,dynes2010,Ast2016,Huang2020,senkpiel2020}, however, the coupling to the electromagnetic environment significantly influences the transport properties \jav{of charged particles}. The electromagnetic field arises from the electric circuit into which the junction is embedded and typically plays an important role in junctions with low capacity (large charging energy). The effects of the coupling between a tunneling particle and the quantized electromagnetic (photonic) bath is conventionally described by the $P(E)$ theory \cite{nazarov, devoret1990, girvin1990, martinis2009, joyezPRL2013}. The theory is based on Fermi's golden rule describing the transmission probability rate between an initial and a final state with equal energies. By taking into account the energy of both the tunneling particles and the photons, and by averaging over the photonic degrees of freedom, one can derive an analytical expression for the $I(V)$ characteristics. The key quantity appearing in the formula is the $P(E)$ function which describes the probability that the tunneling particle emits a photon of energy $E$ to the environment.

Another very commonly used technique to describe the system-bath interaction is the Lindbladian formalism  \cite{Lindblad1976, moy1999, breuer, noneqmethods, Manzano2020, abbruzio2021} which appears quite different in form and spirit:
while the $P(E)$ theory describes inelastic tunneling of particles, the more general Lindblad \jav{master} equation describes the dynamics of open quantum systems.
\jav{Tunneling phenomena have been extensively studied within the framework of quantum master equations \cite{harbolaPRB2006,timm2008,espositoRMP2009,silaev2014,Cuetara_2015}. The typical approach involves tracing out electronic degrees of freedom of the leads and the master equation is formulated for the degrees of freedom of an embedded system, such as a quantum dot.}

In this paper, we present \jav{a different kind of} Lindbladian approach. \jav{We integrate out only} the photonic degrees of freedom, \jav{but keep all electronic degrees of freedom (including those in the leads) as system variables}. We demonstrate how the resulting Lindblad equation \jav{is} used to calculate physical observables.
We emphasize that the Lindbladian approach does not supersede the $P(E)$ approach; the aim of the present work is rather to \jav{compare the conventional $P(E)$ theory to the Lindbladian approach and to demonstrate some benefits of the latter. }
In particular, we compare the assumptions \jav{behind} the conventional Fermi's golden rule and the Lindbladian approach. \jav{The relation between Fermi's golden rule and classical Markovian master equations  has been previously studied in the strict limit of infinitesimally small coupling in Ref.~\onlinecite{Alicki1977}. In this work, we investigate what assumptions apply for small but finite coupling and show that these are essentially the same in both approaches.}
We believe, however, that the Lindbladian approach has one significant technical advantage: \jav{it provides an algorithmic procedure to obtain various quantities related to tunnel phenomena without the necessity of introducing further assumptions. This is demonstrated by considering the problem of Bogoliubov quasi-particle tunneling through superconducting junctions.}

The paper is structured as follows. After defining the problem in Sec.~\ref{sec:problem}, in Sec.~\ref{sec:fermi} we review the main steps of the conventional approach by focusing on the approximations implied in the Fermi's golden rule. These approximations are compared to the assumptions made during the derivation of the Lindbladian equation in Sec.~\ref{sec:lindblad}. In Secs.~\ref{sec:elcurr}, the electric and heat current are calculated using the Lindbladian framework. \jav{In Sec.~\ref{sec:qptunn}, we demonstrate how to apply the formalism to quasiparticle tunneling through superconducting junctions.}
In App.~\ref{sec:dissheat} we provide details on the calculation of the dissipated heat.
In App.~\ref{sec:modbcs} we present the formalism for tracking the number of electrons in superconducting BCS states.

\section{Problem statement}
\label{sec:problem}

\begin{figure}[h]
\centering
\includegraphics[width=8cm]{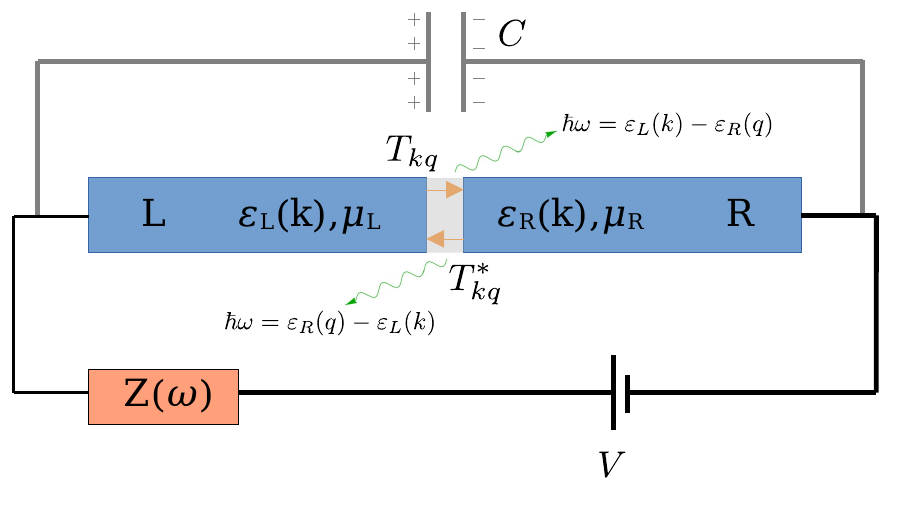}
\caption{Junction embedded into an electric circuit. Orange arrows indicate the possible tunneling processes through the insulating layer (grey). On the two sides of the junction, the left and the right leads are characterized by the energy spectra $\varepsilon_L(k)$ and $\varepsilon_R(q)$ and the chemical potentials of electrons, $\mu_{L/R}$. The electric circuit is characterized by the impedance of $Z(\omega)$, the bias voltage is $V$ and the junction itself has a capacitance $C$.}
\label{fig:setup}
\end{figure}

 The system of interest is the tunnel junction between two leads as shown in Fig.~\ref{fig:setup}. The Hamiltonian of the system is given by
\begin{equation}
H = H_B + H_0 + H_T,
\label{eq:ham}
\end{equation}
\jav{where $H_B$ and $H_T$ are the Hamiltonians describing the electromagnetic bath and the tunneling, while}
\begin{equation}
H_0 = \sum_{k\sigma}\varepsilon_L(k) c_{Lk\sigma}^+ c_{Lk\sigma} + \sum_{q\sigma}\varepsilon_R(q)c_{Rq\sigma}^+ c_{Rq\sigma}
\end{equation}
describes the electrons of the left ($L$) and the right ($R$) leads. Here $c_{Lk\sigma}$ and $c_{Rq\sigma}$ are the annihilation operators of electrons with wavenumber $k/q$ and spin $\sigma$. The \jav{dispersions}  $\varepsilon_L(k)$ and $\varepsilon_R(q)$ \jav{describe the energy spectra of the metallic leads and} will be kept general for most of the calculation.

The leads are coupled to a thermal bath with temperature $T$ and particle reservoirs setting the chemical potentials to $\mu_L$ and $\mu_R$, such that $\mu_L - \mu_R = eV$ with $e$ the charge of the electron and $V$ the bias voltage applied in the electric circuit. 

The tunneling through the junction is described by
\begin{equation}
H_T = \sum_{kq\sigma} \left(T_{kq} c_{Rq\sigma}^+ c_{Lk\sigma} e^{-i\varphi} + h.c.\right), 
\label{eq:singlejunctiontunneling}
\end{equation}
where $e^{-i\varphi}$ is a charge displacement operator acting on the photonic field. From the circuit point of view, the junction is a capacitor. The operator $e^{-i\varphi}$ shifts the charge on the capacitor by the unit charge $e$ \cite{devoret1990,nazarov,martinis2009}. The tunneling amplitudes $T_{kq}$ include the factor of $1/\sqrt{N_L N_R}$ where $N_L$ and $N_R$ designate the number of momentum modes in the left and right lead, respectively.

The term $H_B$ describes the electromagnetic field of the circuit which depends on the actual circuit in which the junction is embedded. 
In the forthcoming discussion, it is assumed that the environment is in thermal equilibrium with the same temperature $T$ as the thermal bath of the leads.

\section{Fermi's golden rule approach}
\label{sec:fermi}

We revisit the conventional derivation of the $P(E)$ theory for a single junction \cite{nazarov}. We will particularly focus on the approximations made in the derivation. These will later be compared to those of the Lindbladian approach.

The $P(E)$ approach is based upon Fermi's golden rule which is used to calculate the transmission rate. \jav{By using first order perturbation theory, }the transmission from the left lead state with energy $\varepsilon_L(k)$ to the right lead state with energy $\varepsilon_R(q)$ is characterized by the \jav{probability} of
\jav{
\begin{eqnarray}
p_{k\sigma,B\rightarrow q\sigma B'}(t) = \frac{\left|\langle q\sigma,B'|H_T|k\sigma,B\rangle\right|^2 }{\hbar^2}\times \nonumber \\
\times \left|\int_0^t e^{\frac{i}{\hbar}\left( \varepsilon_L(k) + E_B - \varepsilon_R(q) - E_{B'}\right)t'}\mathrm{d}t'\right|^2,
\end{eqnarray}
where $B$ and $B'$ denote the bath (photonic) degrees of freedom. In order to compute the total probability of the transition $k\sigma\rightarrow q\sigma$, one has to sum over the bath degrees of freedom:
\begin{equation}
\begin{split}
&p_{k\sigma\rightarrow q\sigma}(t) = \sum_{BB'}\rho_B p_{k\sigma,B\rightarrow q\sigma,B'} \\
&= \frac{|T_{kq}|^2}{\hbar^2}\int_0^t\mathrm{d}t_1 \int_0^t\mathrm{d}t_2 e^{\frac{i}{\hbar}
(\varepsilon_L(k) - \varepsilon_R(q))(t_1 - t_2)} \tilde{P}(t_1-t_2),
\end{split}
\end{equation}
where $\rho_B$ is the probability that the photonic bath is in the state $B$ and
\begin{equation}
\tilde{P}(t) = \Tr_B\left[\hat{\rho}_Be^{i\varphi(t)}e^{-i\varphi(0)}\right]
\end{equation}
is the bath correlation function
with $\hat{\rho}_B$ the density matrix of the electromagnetic environment. It is assumed that $\hat{\rho}_B$ describes a thermal equilibrium state with temperature $T$. In reality, the state of the environment is modified by the coupling but since the transition probabilities are already in second order in the tunneling these modifications result in higher order corrections only. The bath correlation function $\tilde{P}(t)$ can be characterized by several time scales, the longest of which will be denoted by $\tau_B$.

By substituting with the Fourier transform of the bath correlation function,
\begin{equation}
P(E) = \frac{1}{2\pi\hbar}\int_{-\infty}^{\infty}\mathrm{d}t e^{\frac{i}{\hbar}Et}\Tr\left(\hat{\rho}_B e^{i\varphi(t)}e^{-i\varphi(0)}\right),
\label{eq6}
\end{equation}
and integrating over the variables $t_1$ and $t_2$, the transmission probability is rewritten as
\begin{equation}
\begin{split}
&p_{k\sigma\rightarrow q\sigma}(t) = \\
&\frac{|T_{kq}|^2}{\hbar^2}\int_{-\infty}^{\infty}\mathrm{d}E\,P(E) F\left( \frac{\varepsilon_L(k)-\varepsilon_R(q) - E}{\hbar},t\right)
\end{split}
\end{equation}
where
\begin{equation}
F(\omega,t)=\frac{\sin^2\left(\frac{\omega t}{2}\right)}{\left(\frac{\omega}{2}\right)^2}.
\end{equation}
Note that $F(\omega,t)$ is a sharp function around $\omega = 0$ if $t$ is large enough since it has its major contribution in the range of $\left(-\frac{2\pi}{t}; \frac{2\pi}{t}\right)$. If the size of this interval is much smaller than the narrowest energy interval on which $P(E)$ changes significantly in the vicinity of $\varepsilon_L(k) - \varepsilon_R(q)$, the function $F(\omega,t)$ can be replaced by $2\pi t\delta(\omega)$, leading to what is effectively the Fermi's golden rule. The narrowest such energy interval is determined by the longest characteristic time scale $\tau_B$, leading to the condition
$\frac{2\pi}{t}\ll \frac{1}{\tau_B}$
or, dropping the factor of $2\pi$, to
\begin{equation}
\tau_B \ll t.
\end{equation}
If the condition is fulfilled, the transmission probability is obtained as
\begin{equation}
p_{k\sigma\rightarrow q\sigma}(t) = t\frac{2\pi|T_{kq}|^2}{\hbar}P\left(\varepsilon_L(k)-\varepsilon_R(q)\right) \,.
\end{equation}
The probability is valid if it does not reach too high values. By using the notation $R_{kq}=\frac{2\pi|T_{kq}|^2}{\hbar}P\left(\varepsilon_L(k)-\varepsilon_R(q)\right)$, the condition for the validity of the long time limit of the Fermi's golden rule is given by
\begin{equation}
\tau_B\ll t \ll R_{kq}^{-1}.
\label{eq12}
\end{equation}
Such a condition can be fulfilled by some range of $t$ if 
\begin{equation}
\tau_B\ll R_{kq}^{-1}
\label{eq:FGvalid}
\end{equation} 
holds. This means that the Fermi's golden rule is valid if the bath relaxation time is much shorter than the inverse transition rate.

Note that this condition is in spirit the same as presented e.g. in  Chapter 27.6 of Ref.~\cite{zwiebach2022mastering}, but modified due to the averaging over the bath degrees of freedom. If the conditions from Eq.~\eqref{eq12} are fulfilled, the transmission rate for the $k\sigma\rightarrow q\sigma$ transition is finally obtained as
\begin{equation}
w_{k\sigma\rightarrow q\sigma} = \partial_t p_{k\sigma\rightarrow q\sigma}(t) = \frac{2\pi|T_{kq}|^2}{\hbar}P\left(\varepsilon_L(k)-\varepsilon_R(q)\right).
\end{equation}
}

The overall transition rate from left to right is obtained as a sum of the products of the elementary transition probabilities $w_{k\sigma\rightarrow q\sigma}$ and occupation probabilities. 
\jav{Since the transmission rate is already in leading order in the tunneling amplitude, the occupation probabilities in the leads can be assumed to equal their equilibrium value corresponding to}  the temperature $T$ and chemical potentials $\mu_L$ and $\mu_R$.

\jav{Hence, the overall tunneling rate is given by}
\begin{equation}
\overrightarrow{\Lambda} = \frac{2\pi}{\hbar}\sum_{kq\sigma}\left|T_{kq}\right|^2 P(\varepsilon_L(k) - \varepsilon_R(q)) f_L (1- f_R), 
\label{eq:LtoR}
\end{equation}
where we use the shorthand notation $f_L = f(\varepsilon_L(k)- \mu_L)$ and $f_R = f(\varepsilon_R(q)- \mu_R)$ with $f(\xi) = 1/(e^{\beta \xi} + 1)$ the Fermi function and $\beta = 1/(k_{\mathrm{B}} T)$ the inverse temperature. 
The $P(E)$ function is related to the charge relaxation through the electric circuit and can be computed from the properties of the particular circuit \cite{nazarov}. From the electron point of view, the function $P(E)$ represents the probability that the tunneling electron emits a photon with energy $E$ to the environment (for $E>0$) or absorb it (for $E<0$).

Similarly to Eq.~\eqref{eq:LtoR}, the total tunneling rate from right to left can also be obtained and the electric current is finally calculated as $I=e(\overrightarrow{\Lambda} - \overleftarrow{\Lambda})$ \jav{as presented in Ref.~\onlinecite{nazarov}}. 

\section{Lindbladian approach}
\label{sec:lindblad}

In this section, we present an alternative theoretical description of the coupling between the tunneling electrons and the electromagnetic field of the circuit.
We consider the latter as an energy reservoir for the electrons and follow the standard procedure of the microscopic derivation of the Lindblad equation \cite{breuer}. \jav{One} goal of this section is to \jav{compare} the approximations of the Fermi's golden rule to the Born-Markovian approximation used for the Lindblad equation. \jav{The derivation presented here involves tracing out photonic degrees of freedom only and keeping all electronic degrees of freedom as system variables in contrast to many previous theoretical works \cite{harbolaPRB2006,timm2008,espositoRMP2009,silaev2014,Cuetara_2015}.}

In the standard procedure\cite{breuer}, the time evolution is described within the interaction picture with $H_0 + H_B$ being the unperturbed Hamiltonian. The von Neumann equation can be reformulated as
\begin{equation}
\begin{split}
\partial_t \rho = &- \frac{i}{\hbar}\left[H_{TI}(t),\rho(0)\right]  \\
&- \frac{1}{\hbar^2}\int_{0}^{t}\mathrm{d}t'\,\left[H_{TI}(t),\left[H_{TI}(t'),\rho(t')\right]\right],
\end{split}
\label{eq:vN2}
\end{equation}
where $\rho(t)$ is the density matrix of the whole system including both the electronic and photonic degrees of freedom. Square brackets denote the usual commutators and $\rho(0)$ stands for the initial density matrix. $H_{TI}(t)$ denotes the tunneling Hamiltonian in the interaction picture. 

To reduce the Hilbert space to the electronic degrees of freedom, we trace out the environmental degrees of freedom on both sides of the equation. The partial trace leads to $\rho_S(t) = \mathrm{Tr}_B(\rho(t))$ describing solely the electronic degrees of freedom. For the initial condition, we assume that $\mathrm{Tr}_B\left( \left[H_{TI}(t),\rho(0)\right] \right) = 0$. In our specific case of a single junction, this assumption traces back to $\Tr[\rho_B(0) e^{-i\varphi}] = 0$ which holds true if $\rho_B(0)$ describes the thermal equilibrium state of the bath.

Furthermore, it is also assumed that the coupling between the system and the environment is weak, and that the system affects the state of the reservoir only negligibly. Hence, on the right-hand side of the equation, which is already second order in the coupling, we can assume that $\rho(t) = \rho_S(t)\otimes\rho_B$, where $\rho_B$ is the bath density operator in thermal equilibrium with the temperature $T$. Similarly to the Fermi's golden rule approach, the effects of the coupling on the bath would result in higher order corrections only. This assumption is called \textit{Born approximation}.

In the standard procedure, the next assumption is that the environment relaxes much faster than the typical timescales of the system evolution. This \textit{Markovian approximation} can also be rephrased as a statement that the time evolution of the system has no memory effects. Technically, the time argument $t'$ of the density matrix $\rho_S(t')$ is replaced by $t$ indicating that the system density matrix does not change essentially while the bath relaxes. In addition to the change of argument, we also set the lower limit of the integration in Eq.~\eqref{eq:vN2} to $-\infty$ in accordance with the idea that no memory effects from the initial state are kept. \jav{These technical steps lead to the same result as temporal coarse-graining \cite{lidar1999,lidar2001} with the time interval which is much longer than the bath relaxation time $\tau_B$ and is much shorter than the typical timescale of changes in the system.} The physical meaning of the Markovian assumption is that after a tunneling event (also referred to as a jump process in the Lindbladian language) the bath achieves complete relaxation before another event occurs. This is essentially the same assumption as the \jav{long} time limit applied in the Fermi's golden rule approach \jav{as formulated in Eqs. \eqref{eq12} and \eqref{eq:FGvalid}}.

The standard form of the resulting equation is usually achieved by changing the integral variable from $t'$ to $t-t'$ leading to
\begin{eqnarray}
\partial_t\rho_S(t) &= \displaystyle{\frac{1}{\hbar^2}\int_0^\infty} \mathrm{d}t'\,
\mathrm{Tr}_B\Big( H_{TI}(t-t') \rho_S(t)\rho_B H_{TI}(t) -  \nonumber \\
&- H_{TI}(t) H_{TI}(t-t') \rho_S(t)\rho_B + h.c. \Big) \,.
\label{eq:vN3}
\end{eqnarray}

Now, we substitute into this equation the tunneling Hamiltonian which is given by
\begin{equation}
H_{TI}(t) = \sum_p A_p(t)e^{-ij(p)\varphi(t)}
\label{eq:HTI}
\end{equation}
in the interaction picture.
Here, we have defined the process operators $A_p(t) = A_p e^{-i\omega_p t}$ as \jav{shown in Table \ref{tab:normal}}.
\begin{center}
\begin{table}
\def\arraystretch{1.5}
\begin{tabular}{|c|c|c|c|}
\hline
$A_p$ & description & frequency \\
\hline
$T_{kq}c_{q\sigma}^+ c_{k\sigma}$ & $L\rightarrow R$ ($j(p)=+1$) & $\hbar\omega_p = \varepsilon_L(k) - \varepsilon_R(q)$ \\
$T_{kq}^*c_{k\sigma}^+ c_{q\sigma}$ & $R\rightarrow L$ ($j(p)=-1$) & $\hbar\omega_p = \varepsilon_R(q) - \varepsilon_L(k)$ \\
\hline
\end{tabular}
\caption{\jav{Jump processes in the case of normal leads.}}
\label{tab:normal}
\end{table}
\end{center}
\jav{For the jump processes} we have introduced the composite process index $p=\{k,q,\sigma,j\}$. The operator $A_p$ describes a tunneling process between a left lead state with momentum $k$ and a right lead state with momentum $q$, spin $\sigma$ and $j$ is the direction of the tunneling. If the tunneling process occurs from left to right, $j=1$, and $j=-1$ for the opposite direction. In the exponent of Eq.~\eqref{eq:HTI}, $j(p)$ indicates the direction of the process $p$. In the following, we will use $j$ and $j'$ instead of $j(p)$ and $j(p')$ for the sake of brevity.

After substituting into Eq.~\eqref{eq:vN3} and carrying out the integration over $t'$, we obtain
\begin{eqnarray}
\partial_t\rho_S(t) = \frac{1}{\hbar}\sum_{pp'} &\Big[\tilde{\Gamma}_{jj'}\left(\omega_{p'}\right)e^{i(\omega_p - \omega_p')t}\Big( A_{p'}\rho_S(t) A_p^+ - \nonumber \\
&- A_p^+ A_{p'} \rho_S(t)\Big) + h.c.\Big],
\end{eqnarray}
where the bath correlation function has been defined as
\begin{equation}
\tilde{\Gamma}_{jj'}(\omega) = \frac{1}{\hbar}\int_0^\infty \mathrm{d}t\mathrm{Tr}_B\left(\rho_B e^{ij\varphi(t)}e^{-ij'\varphi(0)}\right)e^{i\omega t} \,.
\label{eq:tildegamma}
\end{equation}

In the standard derivation, the exponential factors $e^{i(\omega_p - \omega_{p'})t}$ are assumed to describe very fast oscillations. By applying the rotating wave approximation, we neglect terms where the processes $p$ and $p'$ have different frequencies and keep only terms with $\omega_p = \omega_p'$. In some situations, this approximation does not hold true and one has to handle the nonsecular Lindblad equation with time-dependent coefficients \cite{nonsecularLShnirman,nonsecularLDora}. In the case of tunneling through the single junction, the terms with $\omega_p\neq \omega_{p'}$ do not contribute to the electric current nor to the heat current, hence they would play no role even if retained. 

By applying the rotating wave approximation and returning to the Schr\" odinger picture representation, we obtain
\begin{equation}
\begin{split}
&\partial_t\rho_S(t) = -\frac{i}{\hbar}\left[H_0 + H_{LS},\rho_S(t)\right] +  \\
&+ \frac{1}{\hbar}\sum_{\mathclap{\substack{pp'\\\omega_p = \omega_p'}}}\gamma_{jj'}(\omega_p)
\left(A_{p'} \rho_S(t) A_p^+ - \frac{1}{2}\left\{A_p^+ A_{p'},\rho_S(t)\right\}\right),
\end{split}
\label{eq:lindblad}
\end{equation}
which is the Lindblad equation for the tunneling electrons. 
In this equation, $\{,\}$ denotes anti-commutation. We have also defined 
\begin{equation}
\begin{split}
&\gamma_{jj'}(\omega) = \tilde{\Gamma}_{jj'}(\omega) + \tilde{\Gamma}_{j'j} (\omega)^* =  \\
&=\frac{1}{\hbar}\int_{-\infty}^\infty \mathrm{d}t\mathrm{Tr}_B\left(\rho_B e^{ij\varphi(t)}e^{-ij'\varphi(0)}\right)e^{i\omega t}
\end{split}
\end{equation}
which, apart from a factor of $2\pi$, is the same as the $P(E)$ function from Eq.~\eqref{eq6} if $j=j'$, with $E=\hbar\omega$.
Furthermore, we defined the Lamb shift Hamiltonian as 
\begin{equation}
H_{LS} = \sum_{\mathclap{\substack{pp'\\\omega_p = \omega_p'}}}\frac{1}{2i}\left(\tilde{\Gamma}_{jj'}(\omega) - \tilde{\Gamma}_{j'j}(\omega)^*\right)A_p^+A_{p'} \,.
\end{equation}
So far, we have derived the Lindblad equation for the tunneling particles which is the full dynamical equation for the density matrix.

\section{Electric current}
\label{sec:elcurr}
In this section, we calculate the current flowing through the junction based on the Lindblad equation~\eqref{eq:lindblad}. Since the Lindbladian contains only particle operators, the current must also be expressed in terms of particle operators rather than circuit variables. We start with the expectation value 
\begin{equation}
Q_L(t) = \mathrm{Tr}\left[\hat{Q}_L\rho_S(t)\right]
\end{equation}
where
\begin{equation}
\hat{Q}_L=e\sum_{k\sigma}c_{Lk\sigma}^+ c_{Lk\sigma}
\end{equation}
is the operator for the total charge in the left lead.
The expectation value of the \jav{net} current flowing from left to right is calculated as
\begin{equation}
I=-\frac{\mathrm{d}Q_L}{\mathrm{d}t} = -\mathrm{Tr}\left[\hat{Q}_L\partial_t\rho_S(t)\right] \,.
\end{equation}
We substitute the right-hand side of the Lindblad equation into $\partial_t\rho_S(t)$, make use of the cyclic properties of the trace and the relation $\left[H_0 + H_{LS},c_{Lk\sigma}^+ c_{Lk\sigma} \right] = 0$ which leads to
\begin{equation}
\begin{split}
I = -\frac{1}{\hbar}\sum_{\mathclap{\substack{pp'\\\omega_p = \omega_p'}}}\gamma_{jj'}(\omega_p)\mathrm{Tr}\Bigl[&\rho_S(t)\Bigl(A_{p}^+ \hat{Q}_L A_{p'} \\ 
&- \frac{1}{2}\left\{A_p^+ A_{p'},\hat{Q}_L\right\}\Bigr)\Bigr]\,.
\end{split}
\label{eq:currdef}
\end{equation}

Due to $A_p^{+}A_{p'}\sim T_{kq}^*T_{k'q'}$, the current is already in the order of $|T|^2$ in the tunneling amplitude which is the leading order within the Born approximation. Therefore, we can assume that in the steady state the leads achieve thermal equilibrium, $\rho_S(t\rightarrow\infty) = \rho_\mathrm{th}$. We note that this steady state is not reached by the dissipation described by the jump processes in Eq.~\eqref{eq:lindblad}. In reality, the leads are coupled to other reservoirs: phonons act as a heat bath and wires act as charge reservoirs. The effects of these are expected to be much stronger than the dissipation generated by the tunneling. They thus drive the leads to the thermal equilibrium state with a well-defined temperature and chemical potential.

In thermal equilibrium, $\mathrm{Tr}[\rho_\mathrm{th}c_{Lk\sigma}^+c_{Lk\sigma}]=f_L$ and $\mathrm{Tr}[\rho_\mathrm{th}c_{Rq\sigma}^+c_{Rq\sigma}]=f_R$.
Deviations from these values occur only in higher orders in $|T|^2$. 
When evaluating the trace in Eq.~\eqref{eq:currdef}, terms like $\Tr[\rho_\mathrm{th}A_p^+A_{p'}]$ must be calculated.
Due to the thermal equilibrium, only the $p = p'$ terms contribute
and we obtain
\begin{equation}
\begin{split}
I=\frac{e}{\hbar}\sum_{kq\sigma}|T_{kq}|^2 \left[ \gamma_{++}\left(\frac{\varepsilon_L(k)- \varepsilon_R(q)}{\hbar}\right) f_L\left(1-f_R\right) - \right.  \\
 \left. - \gamma_{--}\left(\frac{\varepsilon_R(q) - \varepsilon_L(k)}{\hbar}\right) f_R \left(1- f_L\right)\right],
 \end{split}
\label{eq:current}
\end{equation}
which is the same as the current obtained from the Fermi's golden rule by evaluating $I=e(\overrightarrow{\Lambda} - \overleftarrow{\Lambda})$ from Eq.~\eqref{eq:LtoR}.
The correspondence is established by
\begin{equation}
\gamma_{++}\left(\frac{E}{\hbar}\right) = \gamma_{--}\left(\frac{E}{\hbar}\right) = 2\pi P(E),
\end{equation}
which directly connects the jump coefficients of the Lindbladian with the $P(E)$ distribution function.

If $T_{kq}$ does not depend on the wavenumber, and the density of states on each lead is also approximated by a constant, $\mathcal{D}_L$ or $\mathcal{D}_R$, the current is calculated as
\begin{eqnarray}
&I=\displaystyle{\frac{1}{eR_T}\int\mathrm{d}\varepsilon_L \int\mathrm{d}\varepsilon_R } \times \\ \times 
&\Big[ P\left(\varepsilon_L- \varepsilon_R\right) f(\varepsilon_L-\mu_L)\left(1-f(\varepsilon_R-\mu_R)\right) -  \nonumber \\
&  - P\left(\varepsilon_R - \varepsilon_L\right) f(\varepsilon_R-\mu_R) \left(1- f(\varepsilon_L-\mu_L)\right)\Big],
\label{eq:current2}
\end{eqnarray}
where
\begin{equation}
R_T = \frac{\hbar}{e^2}\frac{1}{4\pi\mathcal{D}_L\mathcal{D}_R|T|^2}
\label{eq:res}
\end{equation}
is the ohmic resistance of the junction. 
Eq.~\eqref{eq:current2} is identical to Eq.~(51) of Ref.~\cite{nazarov}. 

\jav{
Based on the Lindblad equation, the heat currents can also be computed by taking the time-derivative of the expectation value of the total energy in the left (right) lead $H_L$ ($H_R$), for details see Appendix \ref{sec:dissheat}. Due to the energy exchange between the electrons and the photonic field, the heat current leaving the left lead $P_L$ differs from the heat current arriving to the right lead, $P_R$. The difference is the dissipated heat, $P_{\mathrm{diss}} = P_L - P_R$, which is obtained from the Lindblad equation as
\begin{equation}
\begin{split}
&P_\mathrm{diss} = \frac{2\pi}{\hbar} \sum_{kq\sigma} (\varepsilon_k - \varepsilon_q )|T_{kq}|^2 \times \\ &\times \Big( P(\varepsilon_k - \varepsilon_q) f_L(1-f_R) - P(\varepsilon_q - \varepsilon_k) (1-f_L)f_R\Big).
\end{split}
\end{equation}
which is in accordance with Refs.~\cite{pekolaPRL2007,ruokola2012,rosaDCBthermal2017}. This further confirms the validity of the Lindbladian formalism based on Eq.~\eqref{eq:lindblad}. 
}

In this section, we presented the relation between the conventional approach to the $P(E)$ theory and the Lindbladian formalism \jav{based on tracing out the photonic degrees of freedom}. We have seen that the assumptions of Fermi's golden rule are essentially \jav{the same as} the Born-Markovian approximation used for the Lindblad equation. Namely, both approaches are valid if the coupling between the system and the bath is weak and the results are given up to leading order only. Furthermore, both approaches assume that the tunneling events are independent from each other and that the reservoir completely relaxes between consecutive events.

\jav{
\section{Inelastic quasiparticle tunneling through Josephson junctions}
\label{sec:qptunn}
We apply the Lindbladian formalism to the quasiparticle tunneling between superconducting leads. Note that this phenomenon is distinct from the tunneling of Cooper pairs \cite{nazarov,Alicki_2023} which is not covered in this section. In Ref.~\onlinecite{nazarov}, the quasiparticle tunneling is handled such that the only difference to the normal leads comes from the difference in the density of states of elementary excitations. Ref.~\onlinecite{martinis2009}, however, claims that due to the fact that superconducting elementary excitations are linear superpositions of particles and holes with suitable energy-dependent coefficients, the corresponding coherence factors appear in the expressions for the tunneling rates. Here, we show that by taking into account the Cooper pair counting in the Bogoliubov operators, the results of Ref.~\onlinecite{nazarov} are recovered in special cases but coherence factors appear in more general situations.

Let us start with the Hamiltonian of the superconducting (left) lead which is obtained within mean-field approximation as
\begin{gather}
H_L - \mu_L \hat{N}_L = \sum_{k\sigma} E_L(k) d_{Lk\sigma}^+ d_{Lk\sigma}
\label{eq:hamSCL}
\end{gather}
where $\hat{N}_L$ is the total number of particles, $E_L(k) = \sqrt{\xi_L(k)^2 + |\Delta_L|^2}$ is the energy spectrum with $\Delta_L$ the superconducting gap and $\xi_L(k) = \varepsilon_L(k) - \mu_L$. In the formula, $d_{Lk\sigma}$ is the annihilation operator of the Bogoliubov quasi-particles. Following Refs.~\cite{josephson1962,tinkhamPRB,tinkham}, the annihilation operators are defined as
\begin{gather}
d_{Lk\sigma} = u_k c_{Lk\uparrow} - \sigma v_k S_L c_{L,-k\bar{\sigma}}^+, 
\label{eq:bog}
\end{gather}
where we use the convention that the value of $\sigma$ is +1 for spin up and -1 for spin down and $\bar{\sigma}$ denotes the opposite of spin $\sigma$. Furthermore,
\begin{gather}
u_k = \frac{1}{\sqrt{2}} \sqrt{1 + \frac{\xi_L(k)}{E_L(k)}}\qquad  v_k = \frac{e^{i\phi}}{\sqrt{2}} \sqrt{1 - \frac{\xi_L(k)}{E_L(k)}}
\end{gather}
where $\phi$ is the phase of $\Delta_L$. The non-standard feature of Eq.~\eqref{eq:bog} is the inclusion of the operator $S_L$ which annihilates a Cooper-pair from the the left lead. Technically, we consider that $S_L$ acts on an auxiliary Hilbert space of Cooper-pairs. The basis elements of this Hilbert space are $|M_L\rangle$ describing a state with $M_L$ Cooper pairs and fulfilling $S_L|M_L\rangle = |M_L -1\rangle$. The overall state of the superconducting lead takes the form of a tensor product, $|\mathrm{electrons}\rangle \otimes |M_L\rangle$.

As pointed out in Refs.~\cite{josephson1962,tinkhamPRB,tinkham}, the application of $S_L$ ensures proper accounting for the electric charge, which is necessary for the discussion of tunneling problems. Indeed, as shown in Appendix \ref{sec:modbcs}, the annihilation operators fulfill the relation
\begin{gather}
\left[\hat{N}_{L},d_{Lk\sigma}\right] = - d_{Lk\sigma}
\end{gather}
if the total number of particles operator is defined as 
\begin{gather}
\hat{N}_L = \sum_{k\sigma}c_{Lk\sigma}^+c_{Lk\sigma} + 2\sum_{M_L} M_L |M_L\rangle\langle M_L|,
\label{eq:totnumSC}
\end{gather}
where the factor 2 stems from the fact that Cooper pairs carry a charge of $2e$. 

For the right lead, similar expressions can be derived and with the Hamiltonian $H_R$, the total Hamiltonian of the leads is given as $H_0 = H_L + H_R$. Using the Bogoliubov transformation \eqref{eq:bog}, the tunneling Hamiltonian of \eqref{eq:singlejunctiontunneling} is rewritten as
\begin{widetext}
\begin{gather}
H_T = \sum_{kq\sigma}\Big[T_{kq}e^{-i\varphi}\left( u_k u_q d_{q\sigma}^+ d_{k\sigma} - v_q^* v_k S_R^+ S_L d_{k\sigma}^+ d_{q\sigma}+ \sigma v_q^* u_k S_R^+ d_{q\bar{\sigma}}d_{k\sigma} + \sigma u_q v_k S_L d_{q\sigma}^+ d_{k\bar{\sigma}}^+\right) + \nonumber \\
+ T_{kq}^*e^{i\varphi}\left( u_k u_q d_{k\sigma}^+ d_{q\sigma} - v_q v_k^* S_L^+ S_R d_{q\sigma}^+ d_{k\sigma} + \sigma v_k^* u_q S_L^+ d_{k\bar{\sigma}}d_{q\sigma} + \sigma u_k v_q S_R d_{k\sigma}^+ d_{q\bar{\sigma}}^+\right)\Big],
\end{gather}
from which the jump processes are identified as shown in Table \ref{tab:SC} in accordance with Ref.~\onlinecite{tinkhamPRB}. In the table, the frequencies are determined based on the energy difference between the initial and the final states of the process. For the energy, we consider the eigenvalues of $H_L$ and $H_R$ and not $H_L - \mu_L \hat{N}_L$ or $H_R - \mu_R \hat{N}_R$. That is, we express $H_L$ from Eq.~\eqref{eq:hamSCL} and obtain $H_R$ similarly. This procedure gives rise to the terms $\mu_L - \mu_R = eV$ in the frequencies of the jump processes.
\begin{center}
\begin{table}
\begin{tabular}{|c|c|c|c|}
\hline
id & $A_p$ & description & frequency \\
\hline
1+ & $T_{kq} u_k u_q d_{q\sigma}^+ d_{k\sigma}$ & $L\rightarrow R$, C: - & $\hbar\omega_p = E_k + eV - E_q$ \\
2+ & $-T_{kq} v_q^* v_k S_R^+ S_L d_{k\sigma}^+ d_{q\sigma}$ & $R\rightarrow L$, C: $L\rightarrow R $ & $\hbar\omega_p = E_q + eV - E_k$ \\
3+ & $T_{kq} v_q^* u_k S_R^+ d_{q\bar{\sigma}}d_{k\sigma}$ & $LR\rightarrow \varnothing$, C: $\varnothing\rightarrow R$ & $\hbar\omega_p = E_k + E_q + eV$ \\
4+ & $T_{kq} u_q v_k S_L d_{q\sigma}^+ d_{k\bar{\sigma}}^+$ & $\varnothing\rightarrow LR$, C: $L\rightarrow \varnothing$ & $\hbar\omega_p = eV - E_k - E_q$ \\
\hline
1- & $T_{kq}^* u_k u_q d_{k\sigma}^+ d_{q\sigma}$ & $R\rightarrow L$, C: - & $\hbar\omega_p = E_q - eV - E_k$ \\
2- & $-T_{kq}^* v_q v_k^* S_L^+ S_R d_{q\sigma}^+ d_{k\sigma}$ & $L\rightarrow R$, C: $R\rightarrow L $ & $\hbar\omega_p = E_k - eV - E_q$ \\
3- & $T_{kq}^* u_k v_q S_R d_{k\sigma}^+ d_{q\bar{\sigma}}^+$ & $\varnothing\rightarrow LR$, C: $R\rightarrow \varnothing$ & $\hbar\omega_p = -eV - E_k - E_q$ \\
4- & $T_{kq}^* v_k^* u_q S_L^+ d_{k\bar{\sigma}}d_{q\sigma}$ & $LR\rightarrow \varnothing$, C: $\varnothing\rightarrow L$ & $\hbar\omega_p = E_k + E_q - eV$ \\
\hline
\end{tabular}
\caption{Jump processes in the case of superconducting leads.}
\label{tab:SC}
\end{table}
\end{center}

It is worth to notice the critical importance of the $S_{L,R}$ operators. For example, the processes 1+ and 2- both describe a quasiparticle tunneling from left to right. However, in the process 2-, it is accompanied by a hopping of a Cooper pair from right to left producing a net charge transfer from right to left.

To calculate the electric current through the junction, we follow the same procedure as presented in Sec.~\ref{sec:elcurr}. The total charge of the left lead is given by $\hat{Q}_L = e\hat{N}_L$ with the total number of particles defined in Eq.~\eqref{eq:totnumSC}. Formally, we obtain the same result as Eq.~\eqref{eq:currdef}, but the sum over the different processes includes now all processes of Table~\ref{tab:SC}. We assume again that the density matrix $\rho_S(t\rightarrow\infty)$ describes a thermal equilibrium state of the superconducting leads. As a result, only the diagonal terms $p=p'$ contribute to the sum in Eq.~\eqref{eq:currdef} leading to
\begin{gather}
I = \frac{2\pi e}{\hbar}\sum_{kq\sigma}|T_{kq}|^2\Big[
u_k^2 u_q^2 f_L(1-f_R)P(E_L(k) - E_R(q) + eV) + 
|v_k|^2 |v_q|^2 f_R (1-f_L) P(E_R(q) - E_L(k) + eV) + \nonumber \\  +
u_k^2 |v_q|^2 f_L f_R P(E_L(k) + E_R(q) + eV) +
u_q^2 |v_k|^2 (1-f_L)(1-f_R) P(-E_L(k) - E_R(q) + eV) - \nonumber \\ -
u_k^2 u_q^2 (1-f_L) f_R P(E_R(q) - eV - E_L(k)) -
|v_k|^2 |v_q|^2 f_L (1- f_R) P(E_L(k) - eV - E_R(q)) - \nonumber \\ -
|v_q|^2 u_k^2 (1-f_R) (1- f_L) P(-eV - E_L(k) - E_R(q)) -
u_q^2 |v_k|^2 f_L f_R P(E_L(k) + E_R(q) - eV)
\Big]
\label{eq:currentSC}
\end{gather}
where $f_L = f(E_L(k))$ and $f_R = f(E_R(q))$. 

In the case when both $|T_{kq}|^2$ and the density of states of $\xi_{L/R}$ are independent of the wavenumber and energy, we obtain
\begin{gather}
I = \frac{1}{e R_T}\int_{-\infty}^{\infty}\mathrm{d}\xi_L \int_{-\infty}^{\infty}\mathrm{d}\xi_R \Big[
u_k^2 u_q^2 f_L(1-f_R)P(E_L - E_R + eV) + 
|v_k|^2 |v_q|^2 f_R (1-f_L) P(E_R - E_L + eV) +  \nonumber \\ +
u_k^2 |v_q|^2 f_L f_R P(E_L + E_R + eV) +
u_q^2 |v_k|^2 (1-f_L)(1-f_R) P(-E_L - E_R + eV)- \nonumber \\ -
u_k^2 u_q^2 (1-f_L) f_R P(E_R - eV - E_L) -
|v_k|^2 |v_q|^2 f_L (1- f_R) P(E_L - eV - E_R) - \nonumber \\ -
|v_q|^2 u_k^2 (1-f_R) (1- f_L) P(-eV - E_L - E_R) -
u_q^2 |v_k|^2 f_L f_R P(E_L + E_R - eV)
\Big],
\end{gather}
\end{widetext}
where $E_{L/R}=\sqrt{\xi_{L/R}^2+|\Delta_{L/R}|^2}$ and $R_T$ is the resistance given in Eq.~ \eqref{eq:res}. One can observe that only the coherence factors, e.g., $u_k^2|v_q|^2$, depend on the sign of $\xi_{L/R}$. We analyze all four sectors of the $(\xi_L,\xi_R)$ plane and sum up with respect to the sign of $\xi_{L}$ and $\xi_R$. By using the functions $n_{L/R}(E) = \Theta(|E|-|\Delta_{L/R}|) |E|/\sqrt{E^2 - |\Delta_{L/R}|^2}$ with the Heaviside-function $\Theta(x)$, 
we obtain
\begin{gather}
I=\frac{1}{e R_T}\int_{-\infty}^{\infty}\mathrm{d}E \int_{-\infty}^{\infty}\mathrm{d}E' n_L(E) n_R(E') \times\nonumber \\ \times
\Big[f(E) (1-f(E')) P(E - E' + eV) - \nonumber \\ f(E') (1-f(E)) P(E' - E - eV)\Big],
\end{gather}
which is the same as Eq.~(153) in Ref.~\onlinecite{nazarov}. Note that no coherence factors
appear in this expression.

Let us now study the case of non-constant density of states and assume linear energy dependence in the vicinity of the chemical potential:
\begin{equation}
\begin{split}
\mathcal{D}_L(\xi_k) &= \mathcal{D}_{L0}\left(1 + s_L \xi_k\right), \\
\mathcal{D}_R(\xi_q) &= \mathcal{D}_{R0}\left(1 + s_R \xi_q\right).
\end{split}
\end{equation}
Note that such energy dependence of $\mathcal{D}$ implies that the particle-hole symmetry is broken in the superconducting leads.

By substituting the density of states into \eqref{eq:currentSC} and following the same procedure as previously presented for the constant density case, we now obtain
\begin{gather}
I = \frac{1}{e R_T}\int_{-\infty}^{\infty}\mathrm{d}E \int_{-\infty}^{\infty}\mathrm{d}E' n_L(E) n_R(E') \times \nonumber \\ \times
\left(1 + s_L \frac{E^2 - |\Delta_L|^2}{E}\right) \left(1 + s_R \frac{E'^2 - |\Delta_R|^2}{E'}\right) \times \nonumber \\ \times
\Big[f(E) (1-f(E')) P(E - E' + eV) - \nonumber \\ f(E') (1-f(E)) P(E' - E - eV) \Big],
\label{eq:asym}
\end{gather}
where $R_T$ is defined as Eq.~\eqref{eq:res} but with $\mathcal{D}_{L/R0}$.

The integrals of Eq.~\eqref{eq:asym} are evaluated numerically and the results are shown with solid lines in Fig.~\ref{fig:asym} for an $RLC$ circuit environment where $Q_f=\sqrt{L/C}/R =0.25$ and $L=4C\hbar^2 /e^2$ following Sec. 4.3 in Ref.~\onlinecite{nazarov}.
The numerical calculations have been performed in two distinct cases. In the first (parity symmetric) case, the density of states behave the same in both leads, $s_L = s_R = s$, while in the second (parity anti-symmetric) case, the slopes of the energy dependencies have opposite sign, $s_L = -s_R = s$. The parity anti-symmetric case exhibits more significant $s$-dependence.

At higher bias voltages, the $I(V)$ characteristics converges to the curve corresponding to the case of no environment and no superconductivity. For $s_{L,R}=0$, this reference $I(V)$ function is a linear describing Ohmic behavior. For non-zero $s_{L,R}$, the reference current is calculated as
\begin{equation}
\begin{split}
I_{0}(V) &= \frac{V}{R_T}\Bigl(1 + (s_R - s_L) \frac{e V}{2} \\
&-\frac{s_L s_R}{6}\left((eV)^2 - 2\pi^2 (k_BT)^2 \right)\Bigr)
\label{eq:I0}
\end{split}
\end{equation}
which is plotted with dashed lines in Fig. \ref{fig:asym}. Note that the deviation of the $I(V)$ curves from the Ohmic behaviour at high voltages is not a consequence of the inelastic tunneling but follows from the broken particle-hole symmetry.

The result of Eq.~\eqref{eq:asym} demonstrates how the broken particle-hole symmetry occurs in the $I(V)$ characteristics. The coherence factors $\left(1 + s \frac{E^2 - |\Delta|^2}{E}\right)$ could be derived only by properly identifying the charge conserving jump processes within the Lindbladian formalism.

\begin{figure}[h]
\centering
\includegraphics[width=8cm]{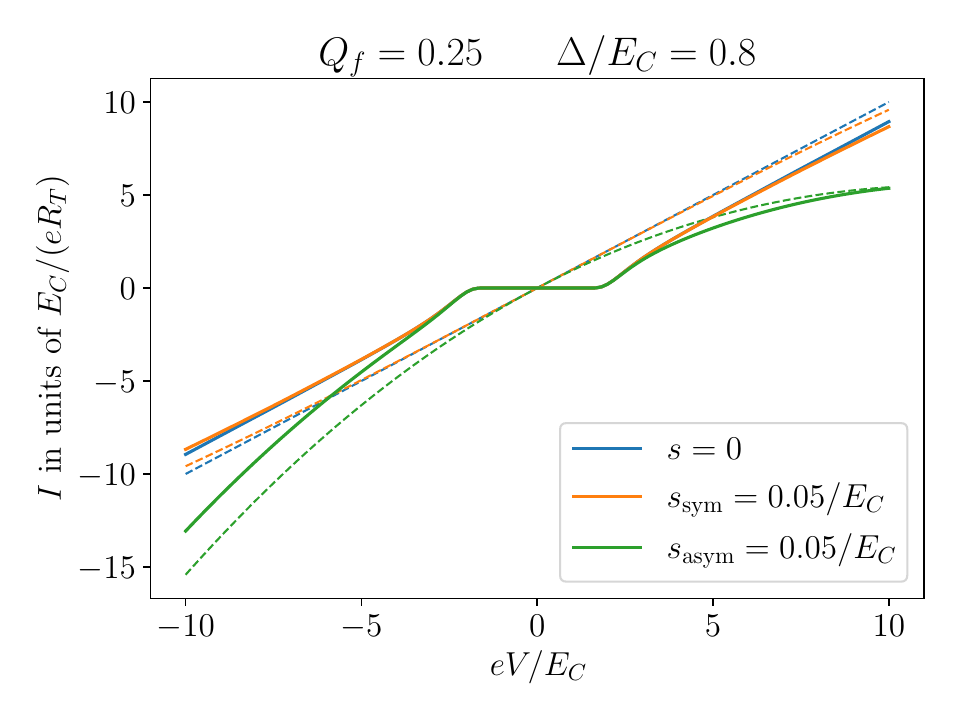}
\caption{$I(V)$ characteristics for different types of density of states. 
The case of uniform density of states (blue, solid) is compared to the parity symmetric case, $s_L = s_R = s_{\mathrm{sym}} = 0.05$ (orange, solid), and to the parity anti-symmetric case, $s_L = s_R = s_{\mathrm{asym}} = 0.05$ (green, solid). 
The dashed lines represent Eq.~\eqref{eq:I0}.}
\label{fig:asym}
\end{figure}

}

\section{Conclusion}
The $P(E)$ theory is the most commonly used approach to explain the transport properties of ultrasmall \jav{tunnel} junctions. 
\jav{In this paper, we showed that the conventional approach to $P(E)$ theory, which is based on Fermi's golden rule, relies on essentially the same assumptions as the Born-Markovian approximation of the Lindblad equation approach. The Lindbladian formalism presented here involves tracing out only photonic degrees of freedom. This has to be contrasted with the typical approaches in the literature where the leads are treated as reservoirs. The main advantage of keeping all electronic degrees of freedom as system variables is that further effects can be included in the model, such as additional dissipative processes or elastic tunneling terms. Furthermore, additional scattering processes could also be taken into consideration. A particularly interesting situation could be to consider an Anderson impurity at the edge of one or both leads and study the interplay between the dissipation and the Kondo effect.}


We believe that our results extend the conventional $P(E)$ theory, open the route to more complex systems, and enable the calculation of more complicated observables. As a demonstration, we have applied the formalism to \jav{inelastic quasiparticle tunneling through superconducting junctions. It has been shown that if the normal-state density of states of the leads is flat (constant in energy), no coherence factors occur in the $I(V)$ characteristics. For a non-constant density of states (i.e., broken particle-hole symmetry), however, the coherence factors play an important role.}

\begin{acknowledgments}
\jav{We thank the anonymous referee for their valuable insights.} We acknowledge the support of the Slovenian Research and Innovation Agency (ARIS) under P1-0416 and J1-3008. \' A.B. acknowledges the support of the National Research, Development and Innovation Office - NKFIH  Project No. K142179.
\end{acknowledgments}

\appendix
\section{Dissipated heat}
\label{sec:dissheat}
 The dissipated heat is a fundamentally important quantity \cite{Lee2013}. Furthermore, it can be used to characterize devices through Joule spectrometry, e.g. in hybrid superconducting devices \cite{joulespec}.
 In this section, we study the heat current flowing through the junction with normal-state leads.
\jav{The heat transport has been broadly studied in the literature \cite{pekolaPRL2007,ruokola2012,rosaDCBthermal2017}. Here, we demonstrate how  these results can be recovered within the Lindbladian formalism.}

 The total energy of the left lead is given by the operator $H_L = \sum_{k\sigma}\varepsilon_k c_{Lk\sigma}^+ c_{Lk\sigma}$ and its expectation value is given by
\begin{equation}
E_L(t) = \Tr\left[H_L \rho_S(t)\right].
\end{equation}
The heat flow from the left lead is computed as
\begin{equation}
P_L = -\frac{\mathrm{d}E_L}{\mathrm{d}t} = -\mathrm{Tr}\left[ H_L\partial_t \rho_S(t)\right].
\end{equation}
By substituting the right-hand side of the Lindblad equation and taking advantage of $\left[H_0 + H_{LS},H_L \right] = 0$, we obtain
\begin{eqnarray}
P_L=\frac{2\pi}{\hbar} \sum_{kq\sigma} \varepsilon_k |T_{kq}|^2 \Big( P(\varepsilon_k - \varepsilon_q) f_L(1-f_R) - \nonumber \\ - P(\varepsilon_q - \varepsilon_k) (1-f_L)f_R\Big),
\end{eqnarray}
where we again \jav{took into account} that the leads are in thermal equilibrium.
Similar formula can be derived for the heat flow to the right lead:
\begin{eqnarray}
P_R=\frac{2\pi}{\hbar} \sum_{kq\sigma} \varepsilon_q |T_{kq}|^2 \Big( P(\varepsilon_k - \varepsilon_q) f_L(1-f_R) - \nonumber \\ - P(\varepsilon_q - \varepsilon_k) (1-f_L)f_R\Big).
\end{eqnarray}
\jav{In the absence of} environmental effects, \jav{when} $P(E) = \delta(E)$, the two heat currents are the same, $P_L = P_R$. In the presence of the photonic bath, however, some part of the heat current which leaves the left lead does not arrive on the right lead but gets dissipated into the environment. The dissipated heat is calculated as
\begin{equation}
\begin{split}
&P_\mathrm{diss} = P_L - P_R = \frac{2\pi}{\hbar} \sum_{kq\sigma} (\varepsilon_k - \varepsilon_q )|T_{kq}|^2 \times \\ &\times \Big( P(\varepsilon_k - \varepsilon_q) f_L(1-f_R) - P(\varepsilon_q - \varepsilon_k) (1-f_L)f_R\Big).
\end{split}
\end{equation}
By considering constant density of states and tunneling amplitude and taking the limits of infinite system size and infinitely large bandwidth, the sum is rewritten as
\begin{equation}
P_\mathrm{diss} = \frac{1-e^{-\beta eV}}{e^2 R_T}\int_{-\infty}^\infty \mathrm{d}E\,\frac{EP(E) (E-eV)}{e^{\beta (E-eV)}-1}.
\end{equation}
In the formula, $R_T$ is the electric resistance of the junction as given in Eq.~\eqref{eq:res}.

\begin{figure}[ht!]
\centering
\includegraphics[width=8.5cm]{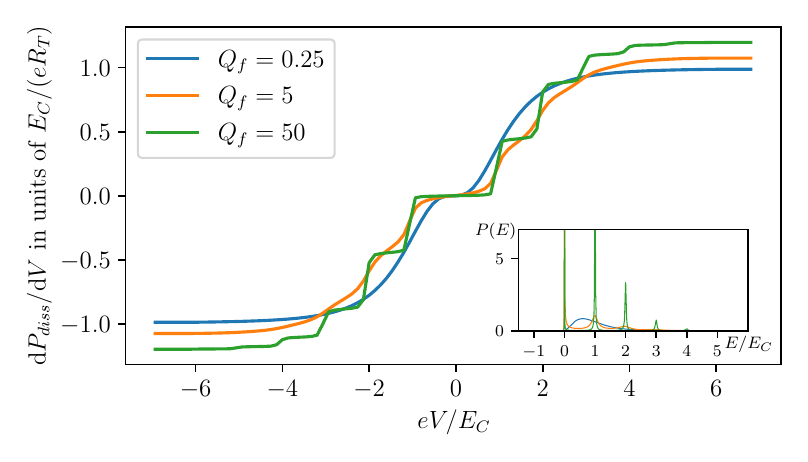}
\caption{Derivative of the dissipated power as a function of bias voltage for three different values of the quality factor $Q_f=\sqrt{L/C}/R$ (numerical results). The inset shows the corresponding $P(E)$ functions in units of $E_C^{-1}$ where $E_C = e^2/(2C)$ is the charging energy.
The temperature is $T=0.01 E_C/k_\mathrm{B}$. The $P(E)$ curves are the same as in Sec. 4.3 of Ref. \cite{nazarov}.
}
\label{fig:dissP}
\end{figure}

The dissipated heat has been computed numerically for an $RLC$ environment, see Fig.~\ref{fig:dissP}.
The circuits are distinguished based on the quality factor defined as $Q_f=\sqrt{L/C}/R$ following Ref.~\cite{nazarov}. In each case, $L=4C\hbar^2 /e^2$ has also been set. The quality factor indicates how pronounced are the resonant peaks in the $P(E)$ function. For lower values of $Q_f$, the ohmic resistance becomes dominant.

At low bias voltage, the dissipated heat has a rich structure and depends strongly on the specific circuit. At high voltages, however, the dissipated heat behaves as $P_{\mathrm{diss}}\sim V$ which is noticeably distinct from the quadratic behavior characteristic of ohmic resistances. This result is particularly interesting since the $I(V)$ characteristics do exhibit the ohmic behavior in this regime, $I=V/R_T$. It can also be shown that  in the large bias voltage limit, the dissipated heat through the junction is given by
\begin{equation}
P_{\mathrm{diss}} = \frac{\bar{E}V}{e R_T},
\end{equation} 
where $\bar{E} = \int\mathrm{d}E EP(E)$ is the average energy emission of a tunneling particle.

We remark that the dissipated heat is not necessarily lost on the ohmic resistance of the electric circuit. It can be shown that even in the case of a simple $LC$ circuit, the dissipated heat is non-zero. This can be explained by \jav{the fact that} the photonic environment with a well-defined temperature is, \jav{by definition}, coupled to a large heat bath and the dissipated heat can also be lost through this coupling.

\jav{
\section{Tracking the number of particles in superconducting leads}
\label{sec:modbcs}
When studying tunneling phenomena \cite{tinkham,tinkhamPRB,josephson1962}, it is beneficial to introduce the Coulomb pair operators $S$ in the annihilation operators of Bogoliubov quasiparticles. 
These operators track the number of Cooper pairs forming the superconducting condensate: $S$ decreases that number by one, $S^+$ increases it by one. The annihilation operators of the Bogoliubov quasiparticles are then given as
\begin{gather}
d_{k\sigma} = u_k c_{k\sigma} - \sigma v_k S c_{-k\bar{\sigma}}^+
\label{eq:bogapp}
\end{gather}
where we use the convention that the value of $\sigma$ is +1 for spin up and -1 for spin down and $\bar{\sigma}$ denotes the opposite of spin $\sigma$.
The coefficients are given by the standard expressions
\begin{equation}
\begin{split}
u_k &= \frac{1}{\sqrt{2}} \sqrt{1 + \frac{\xi_k}{E_k}},\\
v_k &= \frac{e^{i\varphi}}{\sqrt{2}} \sqrt{1 - \frac{\xi_k}{E_k}},
\end{split}
\end{equation}
with $\xi_k = \varepsilon(k) -\mu$ and $E_k=\sqrt{\xi_k^2 + |\Delta|^2}$. The operator $S$ annihilates a Cooper pair from the system, hence, under the action of $d_{k\sigma}$ the charge of the total system decreases by exactly one unit. In this section, we consider only one lead and we have hence dropped the lead index.

The inverse Bogoliubov transformation is given by
\begin{gather}
c_{k\sigma} = u_k d_{k\sigma} + \sigma v_k S d_{-k\bar{\sigma}}^+, 
\label{eq:invbog}
\end{gather}
which is valid if $SS^+ = 1$. This relation will be checked later. An interesting property of Eq.~\eqref{eq:bogapp} is that it does not diagonalize the BCS mean-field Hamiltonian in the standard form
\begin{equation}
\begin{split}
H_{\mathrm{BCS}} &= \sum_{k}\xi_k\left( c_{k\uparrow}^+ c_{k\uparrow} + c_{-k\downarrow}^+ c_{-k\downarrow}\right) \\
&- \sum_{k}\left( \Delta c_{k\uparrow}^+ c_{-k\downarrow}^+ + h.c.\right)
\label{eq:stBCS}
\end{split}
\end{equation}
because the terms of the type $c^+c^+$, which change the total charge by $2e$, cannot be expressed in terms of $d^+ d$, which do not change the total charge. Instead, one has to revisit the mean-field approximation applied to the original Hamiltonian
\begin{equation}
\begin{split}
H &= \sum_{k}\xi_k\left( c_{k\uparrow}^+ c_{k\uparrow} + c_{-k\downarrow}^+ c_{-k\downarrow}\right)\\
&- \frac{g}{N}\sum_{kk'}c_{k\uparrow}^+ c_{-k\downarrow}^+ c_{-k'\downarrow} c_{k'\uparrow}.
\label{eq:ham4}
\end{split}
\end{equation}
Let us insert an identity operator $SS^+$ as
\begin{equation}
\begin{split}
H &= \sum_{k}\xi_k\left( c_{k\uparrow}^+ c_{k\uparrow} + c_{-k\downarrow}^+ c_{-k\downarrow}\right) \\ 
&- \frac{g}{N}\sum_{kk'}c_{k\uparrow}^+ c_{-k\downarrow}^+ SS^+ c_{-k'\downarrow} c_{k'\uparrow}
\label{eq:ham5}
\end{split}
\end{equation}
and assume that the following expectation value
\begin{gather}
\Delta = \frac{g}{N} \sum_{k} \langle S^+ c_{-k\downarrow} c_{k\uparrow} \rangle
\label{eq:gapdef}
\end{gather}
 is non-zero in the ordered phase. By performing the standard steps of the mean-field approximation, we obtain
 \begin{gather}
H_{MF} - \frac{N|\Delta|^2}{g} = \nonumber \\ \sum_{k}\xi_k\left( c_{k\uparrow}^+ c_{k\uparrow} + c_{-k\downarrow}^+ c_{-k\downarrow}\right) - \sum_{k}\left(\Delta c_{k\uparrow}^+ c_{-k\downarrow}^+ S + h.c.\right)
\label{eq:HMF}
\end{gather}
which is readily diagonalized by the transformation \eqref{eq:bogapp} resulting in Eq.~\eqref{eq:hamSCL}.
It is remarkable that, in contrast to \eqref{eq:stBCS}, the Hamiltonian of \eqref{eq:HMF} conserves the total charge because the terms such as $c^+ c^+ S$ describe the creation of two electrons with the simultaneous annihilation of a Cooper pair. Similar description has been recently used in Ref.~\onlinecite{Alicki_2023}.

Technically, by this procedure we have introduced an auxiliary Hilbert space for the Cooper pairs which is separate from the physical electron Hilbert space. This redundancy in description serves to track the number of electrons, which is important in discussing tunneling between multiple superconductors. 
The states of the system take the following general form: 
\begin{equation}
|\Psi\rangle = |\mbox{\{many body state of electrons\}}\rangle \otimes |M\rangle_c
\end{equation}
where the last part, $|M\rangle_c$, describes a state with $M$ Cooper-pairs. In the auxiliary space, the basis vectors are $|M\rangle_c$ with integer $M$ ranging from 0 to infinity. 

The $S$ operator acts as
\begin{equation}
S = \mathrm{I}\otimes \sum_{M=0}^\infty |M\rangle_c \langle M+1|_c
\end{equation}
from which $SS^+ = 1$ also follows. 

The ground state of the Hamiltonian in Eq.~\eqref{eq:HMF} is
\begin{equation}
|\mathrm{GS}\rangle = \prod_{k}\left(u_k + v_k S c_{k\uparrow}^+ c_{-k\downarrow}^+\right) |0\rangle \otimes |M_0\rangle_c,
\end{equation}
where $|0\rangle$ is the vacuum (empty) state of electrons, while $M_0$ is half the total number of electrons in the superconductor. Note that the number of particles is the same in each term of the ground state, $N_e = 2M_0$. Based on the ground state, the gap equation is derived as
\begin{equation}
\Delta = \frac{g}{2N}\sum_k\frac{\Delta}{E_k}
\end{equation}
which coincides with the one in the standard BCS theory \cite{tinkham}.
Finally, we express the operator for the total number of particles by considering both the electron and the Cooper pair contributions as
\begin{equation}
\hat{N} = \sum_{k\sigma}c_{k\sigma}^+c_{k\sigma} + 2\sum_M M |M\rangle\langle M|,
\end{equation}
which satisfies both $[\hat{N},S] = -2S$ and $[\hat{N},d_{k\sigma}] = - d_{k\sigma}$ in accordance with the physical intuition. Note that these commutators are crucial to properly track the charges which is essential when studying transport properties of superconducting junctions.

}

\bibliographystyle{apsrev}
\bibliography{jjdot}

\begin{thebibliography}{51}
\expandafter\ifx\csname natexlab\endcsname\relax\def\natexlab#1{#1}\fi
\expandafter\ifx\csname bibnamefont\endcsname\relax
  \def\bibnamefont#1{#1}\fi
\expandafter\ifx\csname bibfnamefont\endcsname\relax
  \def\bibfnamefont#1{#1}\fi
\expandafter\ifx\csname citenamefont\endcsname\relax
  \def\citenamefont#1{#1}\fi
\expandafter\ifx\csname url\endcsname\relax
  \def\url#1{\texttt{#1}}\fi
\expandafter\ifx\csname urlprefix\endcsname\relax\def\urlprefix{URL }\fi
\providecommand{\bibinfo}[2]{#2}
\providecommand{\eprint}[2][]{\url{#2}}

\bibitem[{\citenamefont{Barone and Paterno}(1982)}]{barone}
\bibinfo{author}{\bibfnamefont{A.}~\bibnamefont{Barone}} \bibnamefont{and}
  \bibinfo{author}{\bibfnamefont{G.}~\bibnamefont{Paterno}},
  \emph{\bibinfo{title}{Physics and applications of the {Josephson} effect}}
  (\bibinfo{publisher}{John Wiley \& Sons}, \bibinfo{year}{1982}).

\bibitem[{\citenamefont{Golubov et~al.}(2004)\citenamefont{Golubov, Kupriyanov,
  and Il'ichev}}]{golubovRMP2004}
\bibinfo{author}{\bibfnamefont{A.~A.} \bibnamefont{Golubov}},
  \bibinfo{author}{\bibfnamefont{M.~Y.} \bibnamefont{Kupriyanov}},
  \bibnamefont{and} \bibinfo{author}{\bibfnamefont{E.}~\bibnamefont{Il'ichev}},
  \bibinfo{journal}{Rev. Mod. Phys.} \textbf{\bibinfo{volume}{76}},
  \bibinfo{pages}{411} (\bibinfo{year}{2004}),
  \urlprefix\url{https://link.aps.org/doi/10.1103/RevModPhys.76.411}.

\bibitem[{\citenamefont{Binnig and Rohrer}(1987)}]{RMPSTM1987}
\bibinfo{author}{\bibfnamefont{G.}~\bibnamefont{Binnig}} \bibnamefont{and}
  \bibinfo{author}{\bibfnamefont{H.}~\bibnamefont{Rohrer}},
  \bibinfo{journal}{Rev. Mod. Phys.} \textbf{\bibinfo{volume}{59}},
  \bibinfo{pages}{615} (\bibinfo{year}{1987}),
  \urlprefix\url{https://link.aps.org/doi/10.1103/RevModPhys.59.615}.

\bibitem[{\citenamefont{Chen}(1993)}]{chen}
\bibinfo{author}{\bibfnamefont{C.~J.} \bibnamefont{Chen}},
  \emph{\bibinfo{title}{Introduction to scanning tunneling microscopy}}
  (\bibinfo{publisher}{Oxford University Press}, \bibinfo{year}{1993}).

\bibitem[{\citenamefont{van Houselt and Zandvliet}(2010)}]{RMPtrSTM2010}
\bibinfo{author}{\bibfnamefont{A.}~\bibnamefont{van Houselt}} \bibnamefont{and}
  \bibinfo{author}{\bibfnamefont{H.~J.~W.} \bibnamefont{Zandvliet}},
  \bibinfo{journal}{Rev. Mod. Phys.} \textbf{\bibinfo{volume}{82}},
  \bibinfo{pages}{1593} (\bibinfo{year}{2010}),
  \urlprefix\url{https://link.aps.org/doi/10.1103/RevModPhys.82.1593}.

\bibitem[{\citenamefont{Ast et~al.}(2016)\citenamefont{Ast, J{\"a}ck, Senkpiel,
  Eltschka, Etzkorn, Ankerhold, and Kern}}]{Ast2016}
\bibinfo{author}{\bibfnamefont{C.~R.} \bibnamefont{Ast}},
  \bibinfo{author}{\bibfnamefont{B.}~\bibnamefont{J{\"a}ck}},
  \bibinfo{author}{\bibfnamefont{J.}~\bibnamefont{Senkpiel}},
  \bibinfo{author}{\bibfnamefont{M.}~\bibnamefont{Eltschka}},
  \bibinfo{author}{\bibfnamefont{M.}~\bibnamefont{Etzkorn}},
  \bibinfo{author}{\bibfnamefont{J.}~\bibnamefont{Ankerhold}},
  \bibnamefont{and} \bibinfo{author}{\bibfnamefont{K.}~\bibnamefont{Kern}},
  \bibinfo{journal}{Nature Communications} \textbf{\bibinfo{volume}{7}},
  \bibinfo{pages}{13009} (\bibinfo{year}{2016}), ISSN
  \bibinfo{issn}{2041-1723},
  \urlprefix\url{https://doi.org/10.1038/ncomms13009}.

\bibitem[{\citenamefont{Karan et~al.}(2022)\citenamefont{Karan, Huang,
  Padurariu, Kubala, Theiler, Black-Schaffer, Morr{\'a}s, Yeyati, Cuevas,
  Ankerhold et~al.}}]{Karan2022}
\bibinfo{author}{\bibfnamefont{S.}~\bibnamefont{Karan}},
  \bibinfo{author}{\bibfnamefont{H.}~\bibnamefont{Huang}},
  \bibinfo{author}{\bibfnamefont{C.}~\bibnamefont{Padurariu}},
  \bibinfo{author}{\bibfnamefont{B.}~\bibnamefont{Kubala}},
  \bibinfo{author}{\bibfnamefont{A.}~\bibnamefont{Theiler}},
  \bibinfo{author}{\bibfnamefont{A.~M.} \bibnamefont{Black-Schaffer}},
  \bibinfo{author}{\bibfnamefont{G.}~\bibnamefont{Morr{\'a}s}},
  \bibinfo{author}{\bibfnamefont{A.~L.} \bibnamefont{Yeyati}},
  \bibinfo{author}{\bibfnamefont{J.~C.} \bibnamefont{Cuevas}},
  \bibinfo{author}{\bibfnamefont{J.}~\bibnamefont{Ankerhold}},
  \bibnamefont{et~al.}, \bibinfo{journal}{Nature Physics}
  \textbf{\bibinfo{volume}{18}}, \bibinfo{pages}{893} (\bibinfo{year}{2022}),
  ISSN \bibinfo{issn}{1745-2481},
  \urlprefix\url{https://doi.org/10.1038/s41567-022-01644-6}.

\bibitem[{\citenamefont{Agra\"it et~al.}(2003)\citenamefont{Agra\"it, Yeyati,
  and van Ruitenbeek}}]{agrait2003}
\bibinfo{author}{\bibfnamefont{N.}~\bibnamefont{Agra\"it}},
  \bibinfo{author}{\bibfnamefont{A.~L.} \bibnamefont{Yeyati}},
  \bibnamefont{and} \bibinfo{author}{\bibfnamefont{J.~M.} \bibnamefont{van
  Ruitenbeek}}, \bibinfo{journal}{Phys. Rep.} \textbf{\bibinfo{volume}{377}},
  \bibinfo{pages}{81} (\bibinfo{year}{2003}).

\bibitem[{\citenamefont{Bretheau et~al.}(2011)\citenamefont{Bretheau, Girit,
  Tosi, Goffman, Joyez, Pothier, Esteve, and Urbina}}]{Bretheau2011}
\bibinfo{author}{\bibfnamefont{L.}~\bibnamefont{Bretheau}},
  \bibinfo{author}{\bibfnamefont{u.}~\bibnamefont{Girit}},
  \bibinfo{author}{\bibfnamefont{L.}~\bibnamefont{Tosi}},
  \bibinfo{author}{\bibfnamefont{M.}~\bibnamefont{Goffman}},
  \bibinfo{author}{\bibfnamefont{P.}~\bibnamefont{Joyez}},
  \bibinfo{author}{\bibfnamefont{H.}~\bibnamefont{Pothier}},
  \bibinfo{author}{\bibfnamefont{D.}~\bibnamefont{Esteve}}, \bibnamefont{and}
  \bibinfo{author}{\bibfnamefont{C.}~\bibnamefont{Urbina}},
  \bibinfo{journal}{Comptes Rendus. Physique} \textbf{\bibinfo{volume}{13}},
  \bibinfo{pages}{89–100} (\bibinfo{year}{2011}), ISSN
  \bibinfo{issn}{1878-1535},
  \urlprefix\url{http://dx.doi.org/10.1016/j.crhy.2011.12.006}.

\bibitem[{\citenamefont{Evers et~al.}(2020)\citenamefont{Evers, Korytár,
  Tewari, and van Ruitenbeek}}]{Evers2020}
\bibinfo{author}{\bibfnamefont{F.}~\bibnamefont{Evers}},
  \bibinfo{author}{\bibfnamefont{R.}~\bibnamefont{Korytár}},
  \bibinfo{author}{\bibfnamefont{S.}~\bibnamefont{Tewari}}, \bibnamefont{and}
  \bibinfo{author}{\bibfnamefont{J.~M.} \bibnamefont{van Ruitenbeek}},
  \bibinfo{journal}{Reviews of Modern Physics} \textbf{\bibinfo{volume}{92}}
  (\bibinfo{year}{2020}), ISSN \bibinfo{issn}{1539-0756},
  \urlprefix\url{http://dx.doi.org/10.1103/RevModPhys.92.035001}.

\bibitem[{\citenamefont{Kouwenhoven and Marcus}(1998)}]{kouwenhoven1998}
\bibinfo{author}{\bibfnamefont{L.}~\bibnamefont{Kouwenhoven}} \bibnamefont{and}
  \bibinfo{author}{\bibfnamefont{C.}~\bibnamefont{Marcus}},
  \bibinfo{journal}{Physics World} \textbf{\bibinfo{volume}{11}},
  \bibinfo{pages}{35} (\bibinfo{year}{1998}).

\bibitem[{\citenamefont{van~der Wiel et~al.}(2003)\citenamefont{van~der Wiel,
  Franceschi, Elzerman, Fujisawa, Tarucha, and Kouwenhoven}}]{wiel2003}
\bibinfo{author}{\bibfnamefont{W.~G.} \bibnamefont{van~der Wiel}},
  \bibinfo{author}{\bibfnamefont{S.~D.} \bibnamefont{Franceschi}},
  \bibinfo{author}{\bibfnamefont{J.~M.} \bibnamefont{Elzerman}},
  \bibinfo{author}{\bibfnamefont{T.}~\bibnamefont{Fujisawa}},
  \bibinfo{author}{\bibfnamefont{S.}~\bibnamefont{Tarucha}}, \bibnamefont{and}
  \bibinfo{author}{\bibfnamefont{L.~P.} \bibnamefont{Kouwenhoven}},
  \bibinfo{journal}{Rev. Mod. Phys.} \textbf{\bibinfo{volume}{75}},
  \bibinfo{pages}{1} (\bibinfo{year}{2003}).

\bibitem[{\citenamefont{Franceschi et~al.}(2010)\citenamefont{Franceschi,
  Kouwenhoven, Sch\"onenberger, and Wernsdorfer}}]{hybrid2010}
\bibinfo{author}{\bibfnamefont{S.~D.} \bibnamefont{Franceschi}},
  \bibinfo{author}{\bibfnamefont{L.}~\bibnamefont{Kouwenhoven}},
  \bibinfo{author}{\bibfnamefont{C.}~\bibnamefont{Sch\"onenberger}},
  \bibnamefont{and}
  \bibinfo{author}{\bibfnamefont{W.}~\bibnamefont{Wernsdorfer}},
  \bibinfo{journal}{Nat. Nanotechnology} \textbf{\bibinfo{volume}{5}},
  \bibinfo{pages}{703} (\bibinfo{year}{2010}).

\bibitem[{\citenamefont{Aiello et~al.}(2022)\citenamefont{Aiello, F{\'e}chant,
  Morvan, Basset, Aprili, Gabelli, and Est{\`e}ve}}]{Aiello2022}
\bibinfo{author}{\bibfnamefont{G.}~\bibnamefont{Aiello}},
  \bibinfo{author}{\bibfnamefont{M.}~\bibnamefont{F{\'e}chant}},
  \bibinfo{author}{\bibfnamefont{A.}~\bibnamefont{Morvan}},
  \bibinfo{author}{\bibfnamefont{J.}~\bibnamefont{Basset}},
  \bibinfo{author}{\bibfnamefont{M.}~\bibnamefont{Aprili}},
  \bibinfo{author}{\bibfnamefont{J.}~\bibnamefont{Gabelli}}, \bibnamefont{and}
  \bibinfo{author}{\bibfnamefont{J.}~\bibnamefont{Est{\`e}ve}},
  \bibinfo{journal}{Nature Communications} \textbf{\bibinfo{volume}{13}},
  \bibinfo{pages}{7146} (\bibinfo{year}{2022}), ISSN \bibinfo{issn}{2041-1723},
  \urlprefix\url{https://doi.org/10.1038/s41467-022-34762-z}.

\bibitem[{\citenamefont{Esaki and Tsu}(1970)}]{5391729}
\bibinfo{author}{\bibfnamefont{L.}~\bibnamefont{Esaki}} \bibnamefont{and}
  \bibinfo{author}{\bibfnamefont{R.}~\bibnamefont{Tsu}}, \bibinfo{journal}{IBM
  Journal of Research and Development} \textbf{\bibinfo{volume}{14}},
  \bibinfo{pages}{61} (\bibinfo{year}{1970}).

\bibitem[{\citenamefont{Sollner et~al.}(1983)\citenamefont{Sollner, Goodhue,
  Tannenwald, Parker, and Peck}}]{Sollner1983}
\bibinfo{author}{\bibfnamefont{T.~C. L.~G.} \bibnamefont{Sollner}},
  \bibinfo{author}{\bibfnamefont{W.~D.} \bibnamefont{Goodhue}},
  \bibinfo{author}{\bibfnamefont{P.~E.} \bibnamefont{Tannenwald}},
  \bibinfo{author}{\bibfnamefont{C.~D.} \bibnamefont{Parker}},
  \bibnamefont{and} \bibinfo{author}{\bibfnamefont{D.~D.} \bibnamefont{Peck}},
  \bibinfo{journal}{Applied Physics Letters} \textbf{\bibinfo{volume}{43}},
  \bibinfo{pages}{588–590} (\bibinfo{year}{1983}), ISSN
  \bibinfo{issn}{1077-3118}, \urlprefix\url{http://dx.doi.org/10.1063/1.94434}.

\bibitem[{\citenamefont{Delsing et~al.}(1989)\citenamefont{Delsing, Likharev,
  Kuzmin, and Claeson}}]{likharev1989}
\bibinfo{author}{\bibfnamefont{P.}~\bibnamefont{Delsing}},
  \bibinfo{author}{\bibfnamefont{K.~K.} \bibnamefont{Likharev}},
  \bibinfo{author}{\bibfnamefont{L.~S.} \bibnamefont{Kuzmin}},
  \bibnamefont{and} \bibinfo{author}{\bibfnamefont{T.}~\bibnamefont{Claeson}},
  \bibinfo{journal}{Phys. Rev. Lett.} \textbf{\bibinfo{volume}{63}},
  \bibinfo{pages}{1180} (\bibinfo{year}{1989}),
  \urlprefix\url{https://link.aps.org/doi/10.1103/PhysRevLett.63.1180}.

\bibitem[{\citenamefont{Pekola et~al.}(2010)\citenamefont{Pekola, Maisi,
  Kafanov, Chekurov, Kemppinen, Pashkin, Saira, M\"ott\"onen, and
  Tsai}}]{dynes2010}
\bibinfo{author}{\bibfnamefont{J.~P.} \bibnamefont{Pekola}},
  \bibinfo{author}{\bibfnamefont{V.~F.} \bibnamefont{Maisi}},
  \bibinfo{author}{\bibfnamefont{S.}~\bibnamefont{Kafanov}},
  \bibinfo{author}{\bibfnamefont{N.}~\bibnamefont{Chekurov}},
  \bibinfo{author}{\bibfnamefont{A.}~\bibnamefont{Kemppinen}},
  \bibinfo{author}{\bibfnamefont{Y.~A.} \bibnamefont{Pashkin}},
  \bibinfo{author}{\bibfnamefont{O.-P.} \bibnamefont{Saira}},
  \bibinfo{author}{\bibfnamefont{M.}~\bibnamefont{M\"ott\"onen}},
  \bibnamefont{and} \bibinfo{author}{\bibfnamefont{J.~S.} \bibnamefont{Tsai}},
  \bibinfo{journal}{Phys. Rev. Lett.} \textbf{\bibinfo{volume}{105}},
  \bibinfo{pages}{026803} (\bibinfo{year}{2010}),
  \urlprefix\url{https://link.aps.org/doi/10.1103/PhysRevLett.105.026803}.

\bibitem[{\citenamefont{Huang et~al.}(2020)\citenamefont{Huang, Padurariu,
  Senkpiel, Drost, Yeyati, Cuevas, Kubala, Ankerhold, Kern, and
  Ast}}]{Huang2020}
\bibinfo{author}{\bibfnamefont{H.}~\bibnamefont{Huang}},
  \bibinfo{author}{\bibfnamefont{C.}~\bibnamefont{Padurariu}},
  \bibinfo{author}{\bibfnamefont{J.}~\bibnamefont{Senkpiel}},
  \bibinfo{author}{\bibfnamefont{R.}~\bibnamefont{Drost}},
  \bibinfo{author}{\bibfnamefont{A.~L.} \bibnamefont{Yeyati}},
  \bibinfo{author}{\bibfnamefont{J.~C.} \bibnamefont{Cuevas}},
  \bibinfo{author}{\bibfnamefont{B.}~\bibnamefont{Kubala}},
  \bibinfo{author}{\bibfnamefont{J.}~\bibnamefont{Ankerhold}},
  \bibinfo{author}{\bibfnamefont{K.}~\bibnamefont{Kern}}, \bibnamefont{and}
  \bibinfo{author}{\bibfnamefont{C.~R.} \bibnamefont{Ast}},
  \bibinfo{journal}{Nature Physics} \textbf{\bibinfo{volume}{16}},
  \bibinfo{pages}{1227} (\bibinfo{year}{2020}), ISSN \bibinfo{issn}{1745-2481},
  \urlprefix\url{https://doi.org/10.1038/s41567-020-0971-0}.

\bibitem[{\citenamefont{Senkpiel et~al.}(2020)\citenamefont{Senkpiel,
  Kl\"ockner, Etzkorn, Dambach, Kubala, Belzig, Yeyati, Cuevas, Pauly,
  Ankerhold et~al.}}]{senkpiel2020}
\bibinfo{author}{\bibfnamefont{J.}~\bibnamefont{Senkpiel}},
  \bibinfo{author}{\bibfnamefont{J.~C.} \bibnamefont{Kl\"ockner}},
  \bibinfo{author}{\bibfnamefont{M.}~\bibnamefont{Etzkorn}},
  \bibinfo{author}{\bibfnamefont{S.}~\bibnamefont{Dambach}},
  \bibinfo{author}{\bibfnamefont{B.}~\bibnamefont{Kubala}},
  \bibinfo{author}{\bibfnamefont{W.}~\bibnamefont{Belzig}},
  \bibinfo{author}{\bibfnamefont{A.~L.} \bibnamefont{Yeyati}},
  \bibinfo{author}{\bibfnamefont{J.~C.} \bibnamefont{Cuevas}},
  \bibinfo{author}{\bibfnamefont{F.}~\bibnamefont{Pauly}},
  \bibinfo{author}{\bibfnamefont{J.}~\bibnamefont{Ankerhold}},
  \bibnamefont{et~al.}, \bibinfo{journal}{Phys. Rev. Lett.}
  \textbf{\bibinfo{volume}{124}}, \bibinfo{pages}{156803}
  (\bibinfo{year}{2020}),
  \urlprefix\url{https://link.aps.org/doi/10.1103/PhysRevLett.124.156803}.

\bibitem[{\citenamefont{Ingold and Nazarov}(1992)}]{nazarov}
\bibinfo{author}{\bibfnamefont{G.-L.} \bibnamefont{Ingold}} \bibnamefont{and}
  \bibinfo{author}{\bibfnamefont{Y.~V.} \bibnamefont{Nazarov}},
  \emph{\bibinfo{title}{Charge Tunneling Rates in Ultrasmall Junctions}}
  (\bibinfo{publisher}{Springer US}, \bibinfo{address}{Boston, MA},
  \bibinfo{year}{1992}), pp. \bibinfo{pages}{21--107}, ISBN
  \bibinfo{isbn}{978-1-4757-2166-9},
  \urlprefix\url{https://doi.org/10.1007/978-1-4757-2166-9{\_}2}.

\bibitem[{\citenamefont{Devoret et~al.}(1990)\citenamefont{Devoret, Esteve,
  Grabert, Ingold, Pothier, and Urbina}}]{devoret1990}
\bibinfo{author}{\bibfnamefont{M.~H.} \bibnamefont{Devoret}},
  \bibinfo{author}{\bibfnamefont{D.}~\bibnamefont{Esteve}},
  \bibinfo{author}{\bibfnamefont{H.}~\bibnamefont{Grabert}},
  \bibinfo{author}{\bibfnamefont{G.-L.} \bibnamefont{Ingold}},
  \bibinfo{author}{\bibfnamefont{H.}~\bibnamefont{Pothier}}, \bibnamefont{and}
  \bibinfo{author}{\bibfnamefont{C.}~\bibnamefont{Urbina}},
  \bibinfo{journal}{Phys. Rev. Lett.} \textbf{\bibinfo{volume}{64}},
  \bibinfo{pages}{1824} (\bibinfo{year}{1990}),
  \urlprefix\url{https://link.aps.org/doi/10.1103/PhysRevLett.64.1824}.

\bibitem[{\citenamefont{Girvin et~al.}(1990)\citenamefont{Girvin, Glazman,
  Jonson, Penn, and Stiles}}]{girvin1990}
\bibinfo{author}{\bibfnamefont{S.~M.} \bibnamefont{Girvin}},
  \bibinfo{author}{\bibfnamefont{L.~I.} \bibnamefont{Glazman}},
  \bibinfo{author}{\bibfnamefont{M.}~\bibnamefont{Jonson}},
  \bibinfo{author}{\bibfnamefont{D.~R.} \bibnamefont{Penn}}, \bibnamefont{and}
  \bibinfo{author}{\bibfnamefont{M.~D.} \bibnamefont{Stiles}},
  \bibinfo{journal}{Phys. Rev. Lett.} \textbf{\bibinfo{volume}{64}},
  \bibinfo{pages}{3183} (\bibinfo{year}{1990}),
  \urlprefix\url{https://link.aps.org/doi/10.1103/PhysRevLett.64.3183}.

\bibitem[{\citenamefont{Martinis et~al.}(2009)\citenamefont{Martinis, Ansmann,
  and Aumentado}}]{martinis2009}
\bibinfo{author}{\bibfnamefont{J.~M.} \bibnamefont{Martinis}},
  \bibinfo{author}{\bibfnamefont{M.}~\bibnamefont{Ansmann}}, \bibnamefont{and}
  \bibinfo{author}{\bibfnamefont{J.}~\bibnamefont{Aumentado}},
  \bibinfo{journal}{Phys. Rev. Lett.} \textbf{\bibinfo{volume}{103}},
  \bibinfo{pages}{097002} (\bibinfo{year}{2009}),
  \urlprefix\url{https://link.aps.org/doi/10.1103/PhysRevLett.103.097002}.

\bibitem[{\citenamefont{Joyez}(2013)}]{joyezPRL2013}
\bibinfo{author}{\bibfnamefont{P.}~\bibnamefont{Joyez}},
  \bibinfo{journal}{Phys. Rev. Lett.} \textbf{\bibinfo{volume}{110}},
  \bibinfo{pages}{217003} (\bibinfo{year}{2013}),
  \urlprefix\url{https://link.aps.org/doi/10.1103/PhysRevLett.110.217003}.

\bibitem[{\citenamefont{Lindblad}(1976)}]{Lindblad1976}
\bibinfo{author}{\bibfnamefont{G.}~\bibnamefont{Lindblad}},
  \bibinfo{journal}{Communications in Mathematical Physics}
  \textbf{\bibinfo{volume}{48}}, \bibinfo{pages}{119} (\bibinfo{year}{1976}),
  ISSN \bibinfo{issn}{1432-0916},
  \urlprefix\url{https://doi.org/10.1007/BF01608499}.

\bibitem[{\citenamefont{Moy et~al.}(1999)\citenamefont{Moy, Hope, and
  Savage}}]{moy1999}
\bibinfo{author}{\bibfnamefont{G.~M.} \bibnamefont{Moy}},
  \bibinfo{author}{\bibfnamefont{J.~J.} \bibnamefont{Hope}}, \bibnamefont{and}
  \bibinfo{author}{\bibfnamefont{C.~M.} \bibnamefont{Savage}},
  \bibinfo{journal}{Phys. Rev. A} \textbf{\bibinfo{volume}{59}},
  \bibinfo{pages}{667} (\bibinfo{year}{1999}),
  \urlprefix\url{https://link.aps.org/doi/10.1103/PhysRevA.59.667}.

\bibitem[{\citenamefont{Breuer and Petruccione}(2002)}]{breuer}
\bibinfo{author}{\bibfnamefont{H.}~\bibnamefont{Breuer}} \bibnamefont{and}
  \bibinfo{author}{\bibfnamefont{F.}~\bibnamefont{Petruccione}},
  \emph{\bibinfo{title}{The Theory of Open Quantum Systems}}
  (\bibinfo{publisher}{Oxford University Press}, \bibinfo{year}{2002}), ISBN
  \bibinfo{isbn}{9780198520634},
  \urlprefix\url{https://books.google.hu/books?id=0Yx5VzaMYm8C}.

\bibitem[{\citenamefont{Landi et~al.}(2022)\citenamefont{Landi, Poletti, and
  Schaller}}]{noneqmethods}
\bibinfo{author}{\bibfnamefont{G.~T.} \bibnamefont{Landi}},
  \bibinfo{author}{\bibfnamefont{D.}~\bibnamefont{Poletti}}, \bibnamefont{and}
  \bibinfo{author}{\bibfnamefont{G.}~\bibnamefont{Schaller}},
  \bibinfo{journal}{Rev. Mod. Phys.} \textbf{\bibinfo{volume}{94}},
  \bibinfo{pages}{045006} (\bibinfo{year}{2022}),
  \urlprefix\url{https://link.aps.org/doi/10.1103/RevModPhys.94.045006}.

\bibitem[{\citenamefont{Manzano}(2020)}]{Manzano2020}
\bibinfo{author}{\bibfnamefont{D.}~\bibnamefont{Manzano}},
  \bibinfo{journal}{AIP Advances} \textbf{\bibinfo{volume}{10}},
  \bibinfo{pages}{025106} (\bibinfo{year}{2020}), ISSN
  \bibinfo{issn}{2158-3226},
  \eprint{https://pubs.aip.org/aip/adv/article-pdf/doi/10.1063/1.5115323/12881278/025106\_1\_online.pdf},
  \urlprefix\url{https://doi.org/10.1063/1.5115323}.

\bibitem[{\citenamefont{D'Abbruzzo and Rossini}(2021)}]{abbruzio2021}
\bibinfo{author}{\bibfnamefont{A.}~\bibnamefont{D'Abbruzzo}} \bibnamefont{and}
  \bibinfo{author}{\bibfnamefont{D.}~\bibnamefont{Rossini}},
  \bibinfo{journal}{Phys. Rev. A} \textbf{\bibinfo{volume}{103}},
  \bibinfo{pages}{052209} (\bibinfo{year}{2021}),
  \urlprefix\url{https://link.aps.org/doi/10.1103/PhysRevA.103.052209}.

\bibitem[{\citenamefont{Harbola et~al.}(2006)\citenamefont{Harbola, Esposito,
  and Mukamel}}]{harbolaPRB2006}
\bibinfo{author}{\bibfnamefont{U.}~\bibnamefont{Harbola}},
  \bibinfo{author}{\bibfnamefont{M.}~\bibnamefont{Esposito}}, \bibnamefont{and}
  \bibinfo{author}{\bibfnamefont{S.}~\bibnamefont{Mukamel}},
  \bibinfo{journal}{Phys. Rev. B} \textbf{\bibinfo{volume}{74}},
  \bibinfo{pages}{235309} (\bibinfo{year}{2006}),
  \urlprefix\url{https://link.aps.org/doi/10.1103/PhysRevB.74.235309}.

\bibitem[{\citenamefont{Timm}(2008)}]{timm2008}
\bibinfo{author}{\bibfnamefont{C.}~\bibnamefont{Timm}}, \bibinfo{journal}{Phys.
  Rev. B} \textbf{\bibinfo{volume}{77}}, \bibinfo{pages}{195416}
  (\bibinfo{year}{2008}),
  \urlprefix\url{https://link.aps.org/doi/10.1103/PhysRevB.77.195416}.

\bibitem[{\citenamefont{Esposito et~al.}(2009)\citenamefont{Esposito, Harbola,
  and Mukamel}}]{espositoRMP2009}
\bibinfo{author}{\bibfnamefont{M.}~\bibnamefont{Esposito}},
  \bibinfo{author}{\bibfnamefont{U.}~\bibnamefont{Harbola}}, \bibnamefont{and}
  \bibinfo{author}{\bibfnamefont{S.}~\bibnamefont{Mukamel}},
  \bibinfo{journal}{Rev. Mod. Phys.} \textbf{\bibinfo{volume}{81}},
  \bibinfo{pages}{1665} (\bibinfo{year}{2009}),
  \urlprefix\url{https://link.aps.org/doi/10.1103/RevModPhys.81.1665}.

\bibitem[{\citenamefont{Silaev et~al.}(2014)\citenamefont{Silaev, Heikkil\"a,
  and Virtanen}}]{silaev2014}
\bibinfo{author}{\bibfnamefont{M.}~\bibnamefont{Silaev}},
  \bibinfo{author}{\bibfnamefont{T.~T.} \bibnamefont{Heikkil\"a}},
  \bibnamefont{and} \bibinfo{author}{\bibfnamefont{P.}~\bibnamefont{Virtanen}},
  \bibinfo{journal}{Phys. Rev. E} \textbf{\bibinfo{volume}{90}},
  \bibinfo{pages}{022103} (\bibinfo{year}{2014}),
  \urlprefix\url{https://link.aps.org/doi/10.1103/PhysRevE.90.022103}.

\bibitem[{\citenamefont{Cuetara and Esposito}(2015)}]{Cuetara_2015}
\bibinfo{author}{\bibfnamefont{G.~B.} \bibnamefont{Cuetara}} \bibnamefont{and}
  \bibinfo{author}{\bibfnamefont{M.}~\bibnamefont{Esposito}},
  \bibinfo{journal}{New Journal of Physics} \textbf{\bibinfo{volume}{17}},
  \bibinfo{pages}{095005} (\bibinfo{year}{2015}),
  \urlprefix\url{https://dx.doi.org/10.1088/1367-2630/17/9/095005}.

\bibitem[{\citenamefont{Alicki}(1977)}]{Alicki1977}
\bibinfo{author}{\bibfnamefont{R.}~\bibnamefont{Alicki}},
  \bibinfo{journal}{International Journal of Theoretical Physics}
  \textbf{\bibinfo{volume}{16}}, \bibinfo{pages}{351} (\bibinfo{year}{1977}),
  ISSN \bibinfo{issn}{1572-9575},
  \urlprefix\url{https://doi.org/10.1007/BF01807150}.

\bibitem[{\citenamefont{Zwiebach}(2022)}]{zwiebach2022mastering}
\bibinfo{author}{\bibfnamefont{B.}~\bibnamefont{Zwiebach}},
  \emph{\bibinfo{title}{Mastering Quantum Mechanics: Essentials, Theory, and
  Applications}} (\bibinfo{publisher}{MIT Press}, \bibinfo{year}{2022}), ISBN
  \bibinfo{isbn}{9780262366892},
  \urlprefix\url{https://books.google.si/books?id=O_E3EAAAQBAJ}.

\bibitem[{\citenamefont{Bacon et~al.}(1999)\citenamefont{Bacon, Lidar, and
  Whaley}}]{lidar1999}
\bibinfo{author}{\bibfnamefont{D.}~\bibnamefont{Bacon}},
  \bibinfo{author}{\bibfnamefont{D.~A.} \bibnamefont{Lidar}}, \bibnamefont{and}
  \bibinfo{author}{\bibfnamefont{K.~B.} \bibnamefont{Whaley}},
  \bibinfo{journal}{Phys. Rev. A} \textbf{\bibinfo{volume}{60}},
  \bibinfo{pages}{1944} (\bibinfo{year}{1999}),
  \urlprefix\url{https://link.aps.org/doi/10.1103/PhysRevA.60.1944}.

\bibitem[{\citenamefont{Lidar et~al.}(2001)\citenamefont{Lidar, Bihary, and
  Whaley}}]{lidar2001}
\bibinfo{author}{\bibfnamefont{D.~A.} \bibnamefont{Lidar}},
  \bibinfo{author}{\bibfnamefont{Z.}~\bibnamefont{Bihary}}, \bibnamefont{and}
  \bibinfo{author}{\bibfnamefont{K.}~\bibnamefont{Whaley}},
  \bibinfo{journal}{Chemical Physics} \textbf{\bibinfo{volume}{268}},
  \bibinfo{pages}{35} (\bibinfo{year}{2001}), ISSN \bibinfo{issn}{0301-0104},
  \urlprefix\url{https://www.sciencedirect.com/science/article/pii/S0301010401003305}.

\bibitem[{\citenamefont{Kamleitner and Shnirman}(2011)}]{nonsecularLShnirman}
\bibinfo{author}{\bibfnamefont{I.}~\bibnamefont{Kamleitner}} \bibnamefont{and}
  \bibinfo{author}{\bibfnamefont{A.}~\bibnamefont{Shnirman}},
  \bibinfo{journal}{Phys. Rev. B} \textbf{\bibinfo{volume}{84}},
  \bibinfo{pages}{235140} (\bibinfo{year}{2011}),
  \urlprefix\url{https://link.aps.org/doi/10.1103/PhysRevB.84.235140}.

\bibitem[{\citenamefont{Vajna et~al.}(2016)\citenamefont{Vajna, Horovitz,
  D\'ora, and Zar\'and}}]{nonsecularLDora}
\bibinfo{author}{\bibfnamefont{S.}~\bibnamefont{Vajna}},
  \bibinfo{author}{\bibfnamefont{B.}~\bibnamefont{Horovitz}},
  \bibinfo{author}{\bibfnamefont{B.}~\bibnamefont{D\'ora}}, \bibnamefont{and}
  \bibinfo{author}{\bibfnamefont{G.}~\bibnamefont{Zar\'and}},
  \bibinfo{journal}{Phys. Rev. B} \textbf{\bibinfo{volume}{94}},
  \bibinfo{pages}{115145} (\bibinfo{year}{2016}),
  \urlprefix\url{https://link.aps.org/doi/10.1103/PhysRevB.94.115145}.

\bibitem[{\citenamefont{Pekola and Hekking}(2007)}]{pekolaPRL2007}
\bibinfo{author}{\bibfnamefont{J.~P.} \bibnamefont{Pekola}} \bibnamefont{and}
  \bibinfo{author}{\bibfnamefont{F.~W.~J.} \bibnamefont{Hekking}},
  \bibinfo{journal}{Phys. Rev. Lett.} \textbf{\bibinfo{volume}{98}},
  \bibinfo{pages}{210604} (\bibinfo{year}{2007}),
  \urlprefix\url{https://link.aps.org/doi/10.1103/PhysRevLett.98.210604}.

\bibitem[{\citenamefont{Ruokola and Ojanen}(2012)}]{ruokola2012}
\bibinfo{author}{\bibfnamefont{T.}~\bibnamefont{Ruokola}} \bibnamefont{and}
  \bibinfo{author}{\bibfnamefont{T.}~\bibnamefont{Ojanen}},
  \bibinfo{journal}{Phys. Rev. B} \textbf{\bibinfo{volume}{86}},
  \bibinfo{pages}{035454} (\bibinfo{year}{2012}),
  \urlprefix\url{https://link.aps.org/doi/10.1103/PhysRevB.86.035454}.

\bibitem[{\citenamefont{Rossell\'o et~al.}(2017)\citenamefont{Rossell\'o,
  L\'opez, and S\'anchez}}]{rosaDCBthermal2017}
\bibinfo{author}{\bibfnamefont{G.}~\bibnamefont{Rossell\'o}},
  \bibinfo{author}{\bibfnamefont{R.}~\bibnamefont{L\'opez}}, \bibnamefont{and}
  \bibinfo{author}{\bibfnamefont{R.}~\bibnamefont{S\'anchez}},
  \bibinfo{journal}{Phys. Rev. B} \textbf{\bibinfo{volume}{95}},
  \bibinfo{pages}{235404} (\bibinfo{year}{2017}),
  \urlprefix\url{https://link.aps.org/doi/10.1103/PhysRevB.95.235404}.

\bibitem[{\citenamefont{Alicki et~al.}(2023)\citenamefont{Alicki, Horodecki,
  Jenkins, Łobejko, and Suárez}}]{Alicki_2023}
\bibinfo{author}{\bibfnamefont{R.}~\bibnamefont{Alicki}},
  \bibinfo{author}{\bibfnamefont{M.}~\bibnamefont{Horodecki}},
  \bibinfo{author}{\bibfnamefont{A.}~\bibnamefont{Jenkins}},
  \bibinfo{author}{\bibfnamefont{M.}~\bibnamefont{Łobejko}}, \bibnamefont{and}
  \bibinfo{author}{\bibfnamefont{G.}~\bibnamefont{Suárez}},
  \bibinfo{journal}{New Journal of Physics} \textbf{\bibinfo{volume}{25}},
  \bibinfo{pages}{113013} (\bibinfo{year}{2023}),
  \urlprefix\url{https://dx.doi.org/10.1088/1367-2630/ad06d8}.

\bibitem[{\citenamefont{Josephson}(1962)}]{josephson1962}
\bibinfo{author}{\bibfnamefont{B.}~\bibnamefont{Josephson}},
  \bibinfo{journal}{Physics Letters} \textbf{\bibinfo{volume}{1}},
  \bibinfo{pages}{251} (\bibinfo{year}{1962}), ISSN \bibinfo{issn}{0031-9163},
  \urlprefix\url{https://www.sciencedirect.com/science/article/pii/0031916362913690}.

\bibitem[{\citenamefont{Tinkham}(1972)}]{tinkhamPRB}
\bibinfo{author}{\bibfnamefont{M.}~\bibnamefont{Tinkham}},
  \bibinfo{journal}{Phys. Rev. B} \textbf{\bibinfo{volume}{6}},
  \bibinfo{pages}{1747} (\bibinfo{year}{1972}),
  \urlprefix\url{https://link.aps.org/doi/10.1103/PhysRevB.6.1747}.

\bibitem[{\citenamefont{Tinkham}(2004)}]{tinkham}
\bibinfo{author}{\bibfnamefont{M.}~\bibnamefont{Tinkham}},
  \emph{\bibinfo{title}{Introduction to Superconductivity}}, Dover Books on
  Physics Series (\bibinfo{publisher}{Dover Publications},
  \bibinfo{year}{2004}), ISBN \bibinfo{isbn}{9780486134727},
  \urlprefix\url{https://books.google.si/books?id=VpUk3NfwDIkC}.

\bibitem[{\citenamefont{Lee et~al.}(2013)\citenamefont{Lee, Kim, Jeong, Zotti,
  Pauly, Cuevas, and Reddy}}]{Lee2013}
\bibinfo{author}{\bibfnamefont{W.}~\bibnamefont{Lee}},
  \bibinfo{author}{\bibfnamefont{K.}~\bibnamefont{Kim}},
  \bibinfo{author}{\bibfnamefont{W.}~\bibnamefont{Jeong}},
  \bibinfo{author}{\bibfnamefont{L.~A.} \bibnamefont{Zotti}},
  \bibinfo{author}{\bibfnamefont{F.}~\bibnamefont{Pauly}},
  \bibinfo{author}{\bibfnamefont{J.~C.} \bibnamefont{Cuevas}},
  \bibnamefont{and} \bibinfo{author}{\bibfnamefont{P.}~\bibnamefont{Reddy}},
  \bibinfo{journal}{Nature} \textbf{\bibinfo{volume}{498}},
  \bibinfo{pages}{209} (\bibinfo{year}{2013}), ISSN \bibinfo{issn}{1476-4687},
  \urlprefix\url{https://doi.org/10.1038/nature12183}.

\bibitem[{\citenamefont{Ibabe et~al.}(2023)\citenamefont{Ibabe, G{\'o}mez,
  Steffensen, Kanne, Nyg{\aa}rd, Yeyati, and Lee}}]{joulespec}
\bibinfo{author}{\bibfnamefont{A.}~\bibnamefont{Ibabe}},
  \bibinfo{author}{\bibfnamefont{M.}~\bibnamefont{G{\'o}mez}},
  \bibinfo{author}{\bibfnamefont{G.~O.} \bibnamefont{Steffensen}},
  \bibinfo{author}{\bibfnamefont{T.}~\bibnamefont{Kanne}},
  \bibinfo{author}{\bibfnamefont{J.}~\bibnamefont{Nyg{\aa}rd}},
  \bibinfo{author}{\bibfnamefont{A.~L.} \bibnamefont{Yeyati}},
  \bibnamefont{and} \bibinfo{author}{\bibfnamefont{E.~J.~H.}
  \bibnamefont{Lee}}, \bibinfo{journal}{Nature Communications}
  \textbf{\bibinfo{volume}{14}}, \bibinfo{pages}{2873} (\bibinfo{year}{2023}),
  ISSN \bibinfo{issn}{2041-1723},
  \urlprefix\url{https://doi.org/10.1038/s41467-023-38533-2}.

\end{thebibliography}

\end{document}